# TRADEMARK SEARCH, ARTIFICIAL INTELLIGENCE, AND THE ROLE OF THE PRIVATE SECTOR

*Sonia K. Katyal*† *& Aniket Kesari*††

## ABSTRACT

Almost every industry today is confronting the potential role that artificial intelligence and machine learning can play in its future. While there are many, many studies on the role of AI in marketing to the consumer, there is less discussion of the role of AI in creating and selecting a trademark that is both distinctive, recognizable, and meaningful to the average consumer. As we argue, given that the role of AI is rapidly increasing in trademark search and similarity areas, lawyers and scholars should be apprised of some of the dramatic implications that AI's role can produce.

We begin, mainly, by proposing that AI should be of interest to anyone studying trademarks and the role that they play in economic decision-making. By running a series of empirical experiments regarding search, we show how comparative work can help us to assess the efficacy of various trademark search engines, many of which draw on a variety of machine learning methods. Traditional approaches to trademarks, spearheaded by economic approaches, have focused almost exclusively on consumer-based, demand-side considerations regarding search. Yet, as we show in this paper, these approaches are incomplete because they fail to take into account the substantial costs that are also faced by not just consumers, but trademark applicants as well. In the end, as we show, machine learning techniques will have a transformative effect on the application and interpretation of foundational trademark doctrines, producing significant implications for the trademark ecosystem. In an age where AI will increasingly govern the process of trademark selection, we argue that the classic division between consumers and trademark owners is perhaps deserving of an updated, supply-side framework. As we argue, a new framework is needed—one that reflects that putative trademark owners, too, are also consumers in the trademark selection ecosystem, and that this insight has transformative potential for encouraging both innovation and efficiency.

DOI: https://doi.org/10.15779/Z380V89H87

© 2020 Sonia K. Katyal & Aniket Kesari.

† Haas Distinguished Chair, University of California, Berkeley; Co-Associate Dean for Faculty Research and Development.

†† Postdoctoral Scholar, Social Sciences D-Lab, University of California at Berkeley; JD Candidate, Yale University. Thanks to Bob Cooter, Stacey Dogan, Aaron Edlin, Benoit Fallenius, Jeanne Fromer, Su Li, Trevor Little, Lisa Larrimore Ouellette, Frank Partnoy, Justin McCrary, Steven Davidoff Solomon, Adam Sterling, Jared Elias, James Hicks, Eric Goldman, Amelia Miazad, Tejas Narechania, Prasad Krishnamurthy, Steven Davidoff Solomon, Robert Bartlett, Tabrez Ebrahim, Barton Beebe, Simone Ross, and Glynn Lunney for their comments. This paper benefited from comments received at a conference with the Berkeley Center for Law, Commerce, and the Economy. We especially thank Jim Dempsey for his helpful insights on the project, the editors at BTLJ for their work, and Mehtab Khan for helpful research assistance. This research was supported by a grant from Markify. The authors affirm that they had complete freedom to direct the study, analysis, results, and reporting, without influence, editorial direction, or censorship.

## TABLE OF CONTENTS

## INTRODUCTION

  Almost every industry today is confronting the potential role that artificial intelligence (AI) and machine learning can play in its future. Intellectual Property (IP) and Information Law are no exception. In areas involving IP, many entities are studying the potential effect of descriptive and predictive analytics on its creation, registration, comparison, and litigation. The U.S. Patent and Trademark Office (USPTO) recently solicited public comments on the relationship between AI and IP,[1] held a conference on the subject, and even ran a contest for improving patent search with AI.[2] More recently, several prominent studies have focused on the role that machine learning can play at the USPTO in the process of prosecution.[3]

  In the area of copyright law, scholars and commentators have voiced significant debate over whether AI-created works can be registered, and the role of human oversight in the crafting of authorship.[4] There are fascinating questions about who owns the rights to an AI-generated work. Does the author of a program, the user, or the AI itself possess the intellectual property rights over these types of works? Determining the scope of authorship in an era where machines are increasingly capable of performing human-like tasks is a fascinating area of IP scholarship.[5] Further, it promises to yield rich debates about the limits of property, personhood, and creativity.

---

  1. *See Request for Comments on Intellectual Property Protection for Artificial Intelligence Innovation*, FEDERAL REGISTER (Oct. 30, 2019), https://www.federalregister.gov/documents/2019/10/30/2019-23638/request-for-comments-on-intellectual-property-protection-for-artificial-intelligence-innovation; *see also* Neil Wilkof, *USPTO Conference on Artificial Intelligence and IP: A Report*, THE IPKAT (Mar. 20, 2019), http://ipkitten.blogspot.com/2019/03/uspto-conference-on-artificial.html.

  2. *See USPTO's Challenge to Improve Patent Search with Artificial Intelligence*, GOVTRIBE (last updated Nov. 7, 2018), https://govtribe.com/opportunity/federal-contract-opportunity/uspto-s-challenge-to-improve-patent-search-with-artificial-intelligence-rfiusptoaipatentseach18.

  3. *See generally* Arti K. Rai, *Machine Learning at the Patent Office: Lessons for Patents and Administrative Law*, 104 IOWA L. REV. 2617 (2019). In this paper, we draw on Rai's instructive description of machine learning, which notes that "a distinctive feature of the genre is that the learning algorithm does not represent the decision rule; instead, the algorithm "learns" the decision rules from data known as training data." *Id.* (citing David Lehr & Paul Ohm, *Playing with the Data: What Legal Scholars Should Learn About Machine Learning*, 51 U.C. DAVIS L. REV. 653 (2017) (explaining machine learning processes)); *see also* Tabrez Y. Ebrahim, *Automation & Predictive Analytics in Patent Prosecution: USPTO Implications & Policy*, 35 GA. ST. U.L. REV. 1185 (2019).

  4. For a lengthier discussion of this literature and the relevant questions, see generally Jane C. Ginsburg & Luke Ali Budiardjo, *Authors and Machines*, 34 BERKELEY TECH. L.J. 343 (2019); Shyam Balganesh, *Causing Copyright*, 117 COLUM. L. REV. 1 (2017).

Yet, surprisingly, very little legal scholarship has addressed the potential role for AI in the context of trademarks.[6] For example, in December 2019, the World Intellectual Property Organization (WIPO) Secretariat issued a draft paper on IP and AI, and while it addressed a range of issues involving the administration of IP and other topics relating to patents, copyright, data, design, and capacity building, it did not cover trademarks.[7] Similarly, while there are many studies on the role of AI in consumer marketing, there is very little scholarly research on the potential role of AI in the corresponding trademark ecosystem.[8] This absence is surprising, especially considering that business owners continue to emphasize that trademarks are the most important area of IP protection.[9] In the United States, IP-related industries

---

5. For a discussion of the intersection with trademark law and economics, see WORLD INTELLECTUAL PROP. ORG., 2013 WORLD INTELLECTUAL PROPERTY REPORT: BRAND – REPUTATION AND IMAGE IN THE GLOBAL MARKETPLACE, 81–108 (2013), https://www.wipo.int/edocs/pubdocs/en/wipo_pub_944_2013-chapter2.pdf.

6. There are very few law-related papers addressing trademarks and AI at the time of publication. *See, e.g.*, Dev Gangjee, *Eye, Robot: Artificial Intelligence and Trade Mark Registers*, *in* TRANSITION AND COHERENCE IN INTELLECTUAL PROPERTY LAW (N. Bruun, G. Dinwoodie, M. Levin & A. Ohly eds., forthcoming 2020), https://papers.ssrn.com/sol3/papers.cfm?abstract_id=3467627; Anke Moerland & Conrado Freitas, Artificial Intelligence and Trade Mark Assessment, *in* Artificial Intelligence & Intellectual Property (R. Hilty, K-C. Liu & J-A. Lee eds., forthcoming 2021), https://papers.ssrn.com/sol3/papers.cfm?abstract_id=3683807.

7. *See* WIPO Conversation on Intellectual Property (IP) and Artificial Intelligence (AI): Second Session, WIPO, https://www.wipo.int/meetings/en/details.jsp?meeting_id=55309 (last visited Jan. 22, 2021).

8. *See, e.g.*, Thomas Davenport, Abhijit Guha, Dhruv Grewal & Timna Bressgott, *How Artificial Intelligence Will Change the Future of Marketing*, J. ACAD. MKTG. SCI. (2019), *available for download at* https://ideas.repec.org/a/spr/joamsc/v48y2020i1d10.1007_s11747-019-00696-0.html; Jan Keitzmann, Jeannette Paschen & Emily Treen, *Artificial Intelligence in Advertising: How Marketers Can Leverage Artificial Intelligence Along the Consumer Journey*, 58 J. ADVERT. RES. 263 (2018); Mònica Casabayó, Nuria Agell & Juan Carlos Aguado, *Using AI Techniques in the Grocery Industry: Identifying the Customers Most Likely to Defect*, 14 INT'L REV. RETAIL DISTRIB. & CONSUMER RES. 295 (2007); Ryan Calo, *Digital Market Manipulation*, 82 GEO. WASH. L. REV. 995 (2014) (offering a look into how technology-mediated advertising intersects with behavioral economics).

9. *See Trademarks, Copyright and Patents: Should Business Owners Really Care About IP?*, VARNUM (May 01, 2019), https://www.varnumlaw.com/newsroom-publications-trademarks-copyrights-and-patents-why-business-owners-should-care-about-ip ("A trademark is one of support at least forty-five million U.S. jobs, contributing over thirty-eight percent to U.S. GDP.[10]

In this Article, we seek to remedy the absence of research in this field by studying the impact of AI on private trademark search engines and their economic and legal implications.[11] We begin by proposing, as a general matter, that AI should be of interest to anyone studying trademarks and the role that they play in economic decision-making. AI will fundamentally transform the trademark ecosystem, and the law will need to evolve as a result. The largest set of questions, we predict, emerges from the need for a more sophisticated approach regarding the impact of AI on the private sector of trademark search. As industries increasingly choose to rely on private AI-powered techniques for search, it becomes more and more essential to consider the nature of these technologies and their implications for trademark creation, comparison, and protection.

In turn, we argue that machine learning will have a transformative effect on the application and interpretation of foundational trademark doctrines. Our study focuses on the application of AI to trademark search and how it fits into a broader discussion about how AI will transform the economics of IP. Most traditional analyses of trademarks focus on the clarifying role of trademarks in aiding consumer search and demand for products in the marketplace. However, we believe that AI carries significant potential to affect the registration and quality of trademarks within the trademark ecosystem, thereby making it necessary to consider the effect of AI on trademark supply as well. Recent increases in trademark applications have exacerbated concerns regarding trademark quality; at least one study has observed, ". . . examiners are going through the motions to meet quota numbers and are not actually

---

the most important business assets that a company will ever own because it identifies and distinguishes the company and its products/services in the marketplace from its competitors."); *see also* Darren Heitner, *Why Intellectual Property is Important for Your Business and What You Should be Doing Now to Protect It*, INC.COM (May 31, 2018), https://www.inc.com/darren-heitner/why-intellectual-property-is-important-for-your-business-what-you-should-be-doing-now-to-protect-it.html (discussing the importance of trademarks).

10. Robert Silvers, Sarah Pearce, Brad Newman, John Phillips, Elena Baca, Tom Brown, Scott Flicker, Emily Pidot, Carson Sullivan & Edward George, *Containing Risk and Seizing Opportunity: The In-house Lawyer's Guide to Artificial Intelligence*, PAUL HASTINGS LLP (Mar. 26, 2019), https://www.paulhastings.com/publications-items/details/?id=43b9226d-2334-6428-811c-ff00004cbded.

11. For a good discussion of various issues that have arisen in the recent rise of trademark applications, see *The Pressure of Rising Demand*, WORLD TRADEMARK REV. (July 1, 2016), https://www.worldtrademarkreview.com/governmentpolicy/pressure-rising-demand [hereinafter WTR Report] (noting rise in application filings and describing the role of the private sector).

examining the evidence."[12] Thus, scholars are increasingly paying attention to the possibility that AI can, and should, be used by the government to even the playing field between itself and potential registrants, in order to improve the quality of registered IP.[13] As AI tools proliferate in the private sector, government failure to adapt could exacerbate market inefficiencies stemming from information asymmetries.[14]

Since there are more trademarks than ever, searching them manually carries enormous costs. Private search algorithms reduce these costs by helping individuals traverse massive datasets efficiently, drawing on AI to do so. While a traditional trademark applicant might rely on government-supported techniques, the Trademark Electronic Search System (TESS), for searching confusingly similar marks, it turns out that TESS is often incomplete. Because of these gaps, several private trademark search engines have emerged to supplement TESS, using machine learning to provide more thorough results. However, not all AI-powered searches are created equal, and their efficacy is a key factor in determining whether users avoid the costs associated with a failed search. Each search engine uses its own methods, algorithms, and techniques to return results. These search engines generally aim to provide a user with a more comprehensive list of potential mark conflicts and to recommend whether the user should proceed with their trademark application, among other services.

As we argue in this Article, a high-level study of AI in the trademark search ecosystem offers us several contributions. To explore the intersection between TESS and private search engines, we conducted a series of experiments to compare the performance of AI-powered search engines in identifying potential conflicts under Section 2(d) of the Trademark Act, 15 U.S.C. § 1052(d),[15] which forbids the registration of a trademark that is confusingly similar to an existing registered trademark. By running a series of comparisons

---

12. *See id.* at 3 (quoting a law firm in its survey responses).
13. *See, e.g.*, Ebrahim, *Automation & Predictive Analytics*, *supra* note 3, at 1188–89 (proposing that the magnified information asymmetries between the inventor and patent examiner can be reduced through artificial intelligence technology).
14. *See id.* at 1189, 1211–28.
15. 15 U.S.C. § 1052(d) (2018). The statute states:

   No trademark by which the goods of the applicant may be distinguished from the goods of others shall be refused registration on the principal register on account of its nature unless it . . . [c]onsists of or comprises a mark which so resembles a mark registered in the Patent and Trademark Office, or a mark or trade name previously used in the United States by another and not abandoned, as to be likely, when used on or in connection with the goods of the applicant, to cause confusion, or to cause mistake, or to deceive . . . . *Id.*

regarding search, we can assess the efficacy of various trademark search engines and study how machine learning methods can plausibly alter the landscape, potentially affecting trademark supply and quality.

Rather than focusing solely on the interaction between the consumer and the producer, our initial results suggest that AI can play a formidable role in addressing the cost of search regarding trademark selection, supply, and quality, warranting a greater focus on trademark producers and the registration ecosystem. While machine learning can minimize some preexisting search costs, our work shows that AI also carries the potential to introduce new search costs into the trademark ecosystem as well.

This work also carries implications for the economic literature regarding trademarks. Traditional approaches to trademarks, spearheaded by economic approaches, have focused almost exclusively on the demand-side role of search costs faced by the consumer. Yet we would argue that the economic literature on search costs, while valuable in considering consumer-based concerns, is incomplete in addressing various issues regarding trademark supply and quality. This conventional economic account fails to also consider the substantial search costs that are faced by not just consumers, but trademark applicants and firms as well in the process of trademark selection.

We argue, primarily, that in an age where AI will increasingly govern the process of trademark selection, this classic division between consumers and trademark owners needs updating, one which reflects that trademark applicants *also* function as consumers in the trademark selection ecosystem. In other words, rather than focusing on the relationships between trademark registrants and buyers or end users of products, we might also focus on how AI-powered search engines flip this dynamic and transform trademark applicants into consumers of trademarks as well. This insight, we suggest, has transformative potential for encouraging both innovation and efficiency in the process of trademark registration. In addition, it also suggests the need to study ways to deploy AI to better optimize search functions, thereby affecting trademark quality and the overall ecosystem as a result.

This Article has four parts. Part I outlines the basic contours of the traditional, demand-side approach in the economic literature focusing on consumer search costs in justifying trademark protection. Part II turns to introducing the role of AI in trademark search, explaining the legal and economic significance of a search cost theory that focuses on trademark supply, rather than demand. Part III turns to our empirical investigation, offering a comparison and contrast of various search engines to demonstrate how supply-side search considerations represent an important aspect of trademark theory. Finally, in Part IV we discuss the legal and economic implications of our research, further exploring the potential role of AI in our legal system for trademarks.

## I.  SEARCH COSTS IN TRADEMARK LAW: A VIEW FROM THE CONSUMER

Back in 1961, George Stigler changed the field of consumer-related economics when he set forth a framework to understand the economic role of information in consumer decision-making.[16] "One should hardly have to tell academicians that information is a valuable resource: knowledge is power," he wrote.[17] "And yet it occupies," he wrote, "a slum dwelling in the town of economics."[18] Yet, if we consider the economic implications of the search for information in the market for goods, he predicted, we can better understand how it affects market price.[19]

Stigler's insight—and the resulting body of literature that followed from it—has come to embody the "informative" view of advertising, one of the dominant approaches to an economic study of advertising.[20] Under this view, which originated out of the Chicago school in the 1960s, consumers often encounter search costs that deter them from learning about a product's availability, price, and quality.[21] Yet advertising, economists argue, can reduce the search costs for this information, improving the efficiency of the marketplace.[22] As we show below, this general view has translated into a specific declaration of the economic and informative value of trademarks in this consumer-centric process of decision-making, a factor that lays the groundwork for a deeper examination of the centrality of search costs in the process of trademark selection.

---

16.  *See generally* George Stigler, *The Economics of Information*, 69 J. POL. ECON. 213 (1961); *see also* Cathy Roheim Wessells, *The Economics of Information: Markets for Seafood Attributes*, 17 MARINE RES. ECON. 153, 154–55 (discussing Stigler).
17.  *Stigler*, *supra* note 16, at 213.
18.  *Id.*
19.  *Id.*
20.  *See generally* KYLE BAGWELL, THE ECONOMIC ANALYSIS OF ADVERTISING 6 (2005) (discussing the informative, persuasive, and complementary view of advertising).
21.  *Id.* at 3.
22.  *See generally* William M. Landes & Richard A. Posner, *Trademark Law: An Economic Perspective*, 30 J. L. & ECON. 265 (1987); Nicholas S. Economides, *The Economics of Trademarks*, 78 TRADEMARK REP. 523 (1988).

A. SEARCH, EXPERIENCE, AND CREDENCE ATTRIBUTES IN CONSUMER DECISION-MAKING

Traditional neoclassical economic theory implied that price signals convey all of the information necessary for consumers to make decisions.[23] However, today only a few markets reflect this phenomenon, because not only are most goods heterogeneous (offering a range of product attributes), but some of those attributes are observable, and others are not.[24] As a result, consumers make their decisions in a world of substantial information asymmetry. However, economists explain, advertising (and relatedly, trademarks) can reduce the costs of obtaining that information.[25] In turn, by offering protection to trademarks, the law thus reduces the search costs consumers face.

By reframing consumer decision-making to include a focus on the willingness to pay for information and the costs of obtaining it, Stigler opened up a world of greater inquiry on how producers communicate information to the public, and the implications of the cost of that information. Years later, in an influential set of papers, Philip Nelson refined Stigler's pathbreaking work by pointing out that there were even greater difficulties associated with ascertaining product quality than price, since information about quality is often impossible to discover before purchase.[26] This view of the consumer's asymmetric search for information has led to the classification of search and experience goods, a framework that underscores the function of trademarks in each category of the marketplace.[27] Others, including Ariel Katz, have since

---

23. *See generally* Jie "Jennifer" Zhang, Xiao Fang & Olivia R. Liu Sheng, *Online Consumer Search Depth: Theories and New Findings*, 23 J. MGMT. INFO. SYS., 72 (2006) ("Existing economic theory modeled consumers' search behavior as a compromise of the anticipated utility gain through price reduction and the additional search cost. Those models assumed that consumers are only searching for a single attribute (e.g., price).").
24. *See generally id.*
25. *Id.* at 82–83 (citing George A. Akerlof, *The Market for "Lemons": Quality Uncertainty and the Market Mechanism*, 84 Q. J. ECON. 488 (1970)); *see also* Landes & Posner, *supra* note 22, at 269 ("Rather than investigating the attributes of all goods to determine which one is brand X or is equivalent to X, the consumer may find it less costly to search by identifying the relevant trademark and purchasing the corresponding brand.").
26. *See* Wessells, *supra* note 16, at 155 (discussing Nelson); *see generally* Phillip Nelson, *Advertising as Information*, 82 J. POL. ECON. 729 (1974) (discussing that there are some qualities of a product which cannot be successfully conveyed by advertising).
27. Phillip Nelson articulated the distinction between search and experience goods; Darby and Karni added a third category, credence goods, to the mix. *See* Phillip Nelson, *Information and Consumer Behavior*, 78 J. POL. ECON. 311, 312 (1970); Michael R. Darby & Edi Karni, *Free Competition and the Optimal Amount of Fraud*, 16 J. L. & ECON. 67, 68–69 (1973).

pointed out that a more precise term might refer to these categories as "attributes," instead of "goods."[28]

Each category nevertheless illustrates the importance of trademarks and advertising in ameliorating the information asymmetry faced by the consumer.[29] Search attributes are qualities that have characteristics which are observable to the consumer, and the brand or producer matters less because the product is readily identifiable (like, for example, table salt).[30] However, in the context of experience attributes, quality can only be determined after consumption of the good, like a newspaper or a law review article that needs to be read first for a consumer to determine its quality.[31] Advertising and trademarks can improve the market for both search and experience attributes because they can provide consumers with pre-purchase information about both price and quality. This, in turn, has the effect of lowering consumers' search costs in reaching decisions.[32]

Later, economists added credence attributes as a third category. These involve goods like pharmaceuticals or automobile repair, where the quality cannot be determined until long after the good has been purchased and consumed.[33] Compared to search attributes and experience attributes, credence attributes are often infeasible to judge even right after purchase, and may take more time to ascertain their quality.[34] Thus, labeling and disclosure-related information can transform a credence attribute into a search attribute in order to empower a consumer to judge the quality of a good prior to making a purchase.[35] Consumers of such goods may perhaps also be aided by a certification of a good by an external source.[36] "For credence goods," Cathy Wessells writes, "one may rely on producer claims, but generally consumers place more trust in an independent third party to provide truthful information on quality," suggesting a role for independent third-party private certification (i.e., certification trademarks) or government regulation.[37]

---

28. Ariel Katz, *Beyond Search Costs: The Linguistic and Trust Functions of Trademarks*, 2010 BYU L. REV. 1555, 1561. We use the terms interchangeably although we note that Katz is correct that attributes is a more precise formulation.

29. *See id.* at 1560–61. Later, Nelson separated products into two different types: search goods and experience goods. *See* Nelson, *Consumer Behavior*, *supra* note 27 (exploring the ways by which a consumer acquires information about the quality of goods)*; see also* Darby & Karni, *supra* note 27, at 68–72 (discussing the importance of credence attributes in assessing the value of the product); George Akerlof, *The Market for "Lemons": Quality Uncertainty and the Market Mechanism*, Q. J. ECON. 488 (1970).

30. Katz, *supra* note 28, at 1560 ("For most consumers, all salt is equally salty, and as long as the consumer can reliably identify the white crystals as salt, the identity of the manufacturer or the exact brand chosen makes very little difference.").

31. *Id.*; *see* Nelson, *Consumer Behavior*, *supra* note 27, at 312.

32. Wessells, *supra* note 16, at 155 (discussing Nelson).

33. Katz, *supra* note 28, at 1561. As Cathy Wessells pointed out, the markets surrounding credence goods are deeply imperfect. This is for two reasons: (1) because of the asymmetry of knowledge between the producer and the consumer and (2) because it is not practical or often even possible for consumers to assess the quality of the product beforehand (e.g., by performing laboratory tests, etc.). Wessells, *supra* note 16, at 155.

34. Darby & Karni, *supra* note 27, at 69.

Taken together, these categories of goods appeared in a substantial amount of economic and legal literature on the foundational role played by advertising—and trademarks—in addressing consumer decision-making. Trademarks, like other forms of advertising, provide important information to both consumers and other producers about their source.[38] In search and experience goods, advertising minimizes the information asymmetry faced by the consumer, enabling her to process information about the good and to decide whether or not to purchase. As Nicholas Economides explains, "Where experience goods have unobservable differences in quality and/or variety, trademarks enable consumers to choose the product with the desired combination of features and encourage firms to maintain consistent quality and variety standards and to compete over a wide quality and variety spectrum."[39] In other words, trademarks convey valuable information for all three categories of attributes, thus justifying their legal protection.

B. TRADEMARK LAW AND CONSUMER SEARCH COSTS

The above analysis describes the role played by trademarks in identifying each of the three core categories of attributes, thereby reducing consumer search costs. Even the Supreme Court has endorsed the search cost justification for trademark protection.[40] In Qualitex, the Court noted,

---

35. Wessells, *supra* note 16, at 155 (citing Caswell). *See* Julie A. Caswell, *Valuing the Benefits and Costs of Improved Food Safety and Nutrition*, 42 AUSTL. J. AGRIC. & RES. ECON. 409 (1998).
36. Darby & Karni, *supra* note 27, at 69–70 (outlining credence goods, by taking the example of repair services, which basically requires a consumer to purchase both information (about the diagnosis of, say, a malfunctioning machine) and repair (actual performance of the repair)). If there were no additional costs involved in separating the two then the authors suggest that the consumer would do so in order to avoid the possibility of fraud. But since it is often cheaper to provide information and service jointly, then the consumer will purchase them both from the same source.
37. Wessells, *supra* note 16, at 155.
38. Stacey L. Dogan & Mark A. Lemley, *Trademarks and Consumer Search Costs on the Internet*, 41 HOUS. L. REV. 777, 777–78 (2004).
39. Economides, *supra* note 22, at 525.
40. *See* Mark P. McKenna, *A Consumer Decision-Making Theory of Trademark Law*, 98 VA. L. REV. 67, 75–76 (2012) ("The overwhelming majority of scholars use search costs language to describe trademark law's purposes, and the Supreme Court has explicitly endorsed the theory as trademark law's core theoretical justification." (internal citation and quotations omitted));

"[T]rademark law, by preventing others from copying a source-identifying mark, 'reduce[s] the customer's costs of shopping and making purchasing decisions,' for it quickly and easily assures a potential customer that this item—the item with this mark—is made by the same producer as other similarly marked items that he or she liked (or disliked) in the past."[41]

Similarly, William Landes and Richard Posner frame trademarks primarily as an informational mechanism to provide consumers with information about the seller's identity, the quality of the product, etc., and thereby reduce the consumer's search costs for comparable goods.[42] The search cost approach has had multiple implications for trademark law; among them are reinforcing the centrality of the consumer and also indirectly empowering strong marks over weaker ones.[43] As Barton Beebe has pointed out, the more distinctive the mark, the less costly it is for the consumer to locate in the marketplace; thus, stronger marks better facilitate the search process for consumers than weaker marks.[44] Trademarks also help guarantee market quality, ameliorating the market failure George Akerlof identified in his famous piece.[45] Not only do they reduce search costs by condensing complex information into an identifiable symbol, but they also "allow buyers to trust and rely upon the signals conveyed by

---

*see also* WILLIAM M. LANDES & RICHARD A. POSNER, THE ECONOMIC STRUCTURE OF INTELLECTUAL PROPERTY LAW 166–209 (2003); John F. Coverdale, *Trademarks and Generic Words: An Effect-on-Competition Test*, 51 U. CHI. L. REV. 868, 869–70, 878 (1984); Stacey L. Dogan & Mark A. Lemley, *A Search-Costs Theory of Limiting Doctrines in Trademark Law*, 97 TRADEMARK REP. 1223, 1223 (2007); Stacey L. Dogan & Mark A. Lemley, *Grounding Trademark Law Through Trademark Use*, 92 IOWA L. REV. 1669, 1689–90, 1697 (2007); Economides, *supra* note 22, at 525–27; Michael Grynberg, *The Road Not Taken: Initial Interest Confusion, Consumer Search Costs, and the Challenge of the Internet*, 28 SEATTLE U.L. REV. 97, 97–99 (2004); William M. Landes & Richard A. Posner, *The Economics of Trademark Law*, 78 TRADEMARK REP. 267, 272 (1988); Mark A. Lemley, *The Modern Lanham Act and the Death of Common Sense*, 108 YALE L.J. 1687, 1695–96 (1999); Clarisa Long, *Dilution*, 106 COLUM. L. REV. 1029, 1033–34, 1056 (2006); Glynn S. Lunney, Jr., *Trademark Monopolies*, 48 EMORY L.J. 367, 432 (1999); I.P.L. Png & David Reitman, *Why Are Some Products Branded and Others Not?*, 38 J.L. & ECON. 207, 208–11 (1995).

41. *Qualitex Co. v. Jacobson Prods. Co., Inc.*, 514 U.S. 159, 163–64 (1995) (internal citations omitted).

42. *See* Landes & Posner, *Trademark Law: An Economic Perspective*, *supra* note 22, at 269–70.

43. For an excellent account of the multiple roles of search in trademark law, see Barton Beebe, *Search and Persuasion in Trademark Law*, 103 MICH. L. REV. 2020, 2042 (2005).

44. *Id.* at 2042–43.

45. *See* Akerlof, *supra* note 29 (arguing that in situations where sellers and buyers have asymmetric information about the quality of a good (i.e., with a used car), adverse selection will occur where high-quality sellers leave the market as consumer willingness-to-pay falls). To avoid this type of market failure, building credible signals of product quality is crucial, and advertising can help achieve this goal.

sellers as guarantees for quality, thus helping to prevent the lemonization of markets for goods with experience and credence attributes."[46]

Firms that produce experience or credence goods are therefore incentivized to keep a consistent level of quality associated with their goods in order to ensure repeat purchasers; trademarks reduce search costs in both of these arenas, enabling the consumer to trust that the purchase they are making will be consistent with their prior experience.[47] But, as Mark Lemley and Stacey Dogan explain, there is a crucial catch: this only works if consumers can readily trust the information that trademarks provide, thereby paving the way for the role of law.[48] "By protecting established trademarks against confusing imitations," they write, "the law ensures a reliable vocabulary . . . . Both sellers and buyers benefit from the ability to trust this vocabulary to mean what it says it means."[49] Because trademarks economize on information, it is thought that making it less costly to obtain will better inform consumers and thereby improve the competitiveness of the market.[50]

Despite the potentially rich layers of focus on trademark owners and applicants for discussion, no other theory has managed to displace the primary importance of the search-cost rationale and its consumer-centric focus. Mark McKenna has valuably pointed out that trademark law itself predated the search cost theory by several hundred years, suggesting that a historical account might be a better, more comprehensive theory to address its development.[51] Other scholars have written about how trademark protection performs a "signaling" function within advertising; others have focused on how brands facilitate corporate growth into new territories; and still others focus on how trademarks are viewed as a kind of property right.[52] Yet, despite

---

46. Katz, *supra* note 28, at 1563.
47. Katz, *supra* note 28, at 1561. While these classes of goods are incredibly helpful in distilling the marketplace, Ariel Katz reminds us that in more contemporary parlance, it is more correct to refer to attributes instead of goods.

> For example, the fact that a can of tuna looks like a can of tuna is a search attribute. The fact that the content tastes like tuna is an experience attribute. Whether the content is indeed tuna and not a good imitation, or whether it is safe for consumption, are credence attributes.

*Id.* at 1561.
48. Stacey L. Dogan & Mark A. Lemley, *Trademarks and Consumer Search Costs on the Internet*, 41 HOUS. L. REV. 777, 786–87 (2004).
49. *Id.* at 787.
50. *Id.*
51. McKenna, *supra* note 40, at 67.
52. Dogan & Lemley, *Trademarks and Consumer Search Costs on the Internet*, *supra* note 48, at 799; *see also* Ralph S. Brown Jr., *Advertising and the Public Interest: Legal Protection of Trade Symbols*, 57 Yale L.J. 1165, 1184 (1948); Lemley, *The Modern Lanham Act and the Death of Common Sense*, the promise of these alternative approaches, search cost theory still plays a seminal role in trademark law, often ensuring the consumer's centrality to trademark law, at times even at the expense of a trademark owner.

Multiple doctrines of trademark law—distinctiveness, genericness, dilution, comparative advertising, and even the theory of trademark use—implicitly follow the search cost approach in crafting legal entitlements.[53] For example, the goal of limiting search costs has been implicitly extended to explain the genericness doctrine, in order to avoid the risk that "[c]onsumers will be misled if what they believe is a generic term is in fact a product sold by only one company."[54] The search cost rationale has also been extended to justify Congress's foray into enacting federal anti-dilution protections, under the reasoning that uses that blur or tarnish famous marks increase the search costs faced by the consumer by either weakening the meaning of the mark in the eyes of the consumer or creating a negative impression of or association with the mark.[55] In sum, trademarks have served as a vehicle to optimize consumer access to information through reducing search costs, and much of trademark law has integrated this goal throughout various doctrines.

## II. SEARCH COSTS IN TRADEMARK REGISTRATION: A VIEW FROM A TRADEMARK APPLICANT

As we discussed above, the conventional legal accounts of search costs focus largely on improving the information shared with the consumer. But this view can often be too narrow. Very little attention is paid to the process of optimizing the information markets that develop around the process of trademark search and registration, even though these variables can have a dramatic effect on trademark supply and enforcement.[56] However justifiable

---

*supra* note 40, at 1714; Kenneth L. Port, *Trademark Monopolies in the Blue Nowhere*, 28 WM. MITCHELL L. REV. 1091 (2002); Lunney, Jr., *Trademark Monopolies*, *supra* note 40; Frank I. Schechter, *Fog and Fiction in Trade-Mark Protection*, 36 COLUM. L. REV. 60, 65 (1936).

53. *See* Dogan & Lemley, *supra* note 48, at 786–99.

54. At the same time, however, Lemley and Dogan point out that the genericness doctrine can actually increase search costs if an ultra-famous mark like "aspirin" or "thermos" has now become generic, since consumers who might associate the mark with a particular source may now be confused if the term is used to refer to a class of goods instead. *See id.* at 793.

55. *Id.* at 789–90*; see also* Rebecca Tushnet, *Gone in Sixty Milliseconds: Trademark Law and Cognitive Science*, 86 TEX. L. REV. 507 (2008) (noting the argument, aided by cognitive science, that negative trademarks (either ones that weaken or tarnish a mark) can create informational harms that reduce consumers' capacity to shop around in a rational manner).

56. Of course, see the seminal paper by Beebe and Fromer, which valuably focused on the issue of trademark supply. *See* Barton Beebe & Jeanne C. Fromer, *Are We Running Out of* the search cost approach may be, it can affect the trademark supply if it adds too much strength to established marks at the cost of others. Too much empowerment of trademark holders can enable them to exert overbroad control over uses that may not even be legitimate trademark uses, or to stifle competitors who are simply describing their own products.[57] As Lemley and Dogan point out, stronger trademark entitlements can also have the effect of narrowing the scope of available words for others to use.[58]

Moreover, despite all of the analysis surrounding the consumer, there is very little recognition of the fact that trademark registrants are also consumers as well in the marketplace of trademark search and registration. Even aside from the law's role in registration, the selection of a trademark is a crucial moment for a firm because it symbolizes much more than the source of the product itself. Since the goal of modern marketing and branding is to essentially create desire among consumers by making irrelevant attributes seem relevant and valuable,[59] the selection of an appropriate trademark is an emotionally-driven choice as well as an economic one.[60] Brands confer market power. As one author writes, "when trademarks protect brands with significant image value, the brand in and of itself becomes a product characteristic that consumers care about but competitors cannot copy."[61]

Thus, the same price and non-price variables that might influence a consumer's purchasing decision might also influence a trademark registrant's decision to select a mark. Even information about the demographics of the typical and non-typical trademark registrants and their trademark search processes or sophistication with online search would be enormously helpful in future research.[62] AI-driven tools could play a crucial role in this process at all levels ranging from trademark selection, to application, and to registration.

Moreover, in a world characterized by more trademarks than ever, it becomes necessary to explore the costs incurred by firms themselves in the process of searching for available trademarks. Trademark applicants will

---

*Trademarks? An Empirical Study of Trademark Depletion and Congestion*, 131 HARV. L. REV. 945, 947 (2018).

57. Dogan & Lemley, *supra* note 48, at 788.
58. *Id.*
59. *See* McKenna, *supra* note 40, at 115 (citing Gregory S. Carpenter et al., *Meaningful Brands from Meaningless Differentiation: The Dependence on Irrelevant Attributes*, 31 J. MKTG. RES. 339, 339 (1994)).
60. WORLD INTELLECTUAL PROP. ORG., *supra* note 5, at 86. *See generally* Sonia Katyal, *Stealth Marketing and Antibranding: The Love that Dare not Speak its Name*, 58 BUFF. L. REV. 58 (2010) (discussing the lure of branding); Sonia Katyal, *Trademark Cosmopolitanism*, 47 UC DAVIS L. REV. 875 (2013) (discussing the emergence of brands as global figures of speech).
61. WORLD INTELL. PROP. ORG., *supra* note 5, at 86.
62. *See* Zhang et al., *supra* note 23, at 91(noting the role of similar attributes for a typical study of consumer search behavior).

expend tremendous effort and incur costs in order to find their optimal trademark for both economic and non-economic reasons. These kinds of search costs seem to be underexplored in the relevant trademark literature, but they are important. Because of the economic benefits of maintaining trustworthy trademarks, the USPTO will reject trademark applications that risk trademark infringement or dilution. To avoid this risk, a firm will ideally want to avoid the costs associated with filing a doomed application, and instead preemptively search for existing marks and calculate the probability of infringement or dilution based on those search results. For this reason, AI and machine learning can play a significant role in improving trademark quality and registrability, reducing the search costs faced by trademark applicants.[63]

Below, we outline the theoretical basis for studying how private AI-powered search tools have emerged to play an important role in supplementing government determinations and reducing search costs faced by the trademark applicant. We then turn to the specifics of discussing how AI is used by government agencies in administering IP and by private entities in the process of search, registration, and brand management.

### A. SUPPLEMENTING TRADEMARK SEARCH IN THE PRIVATE SECTOR

In Part I, we discussed the traditional economic underpinnings of trademarks from the consumer's point of view. Specifically, we discussed the need for the USPTO to avoid granting marks that would result in informational harms to consumers. An erroneously granted trademark creates harms to consumers by confusing them and eroding their ability to discern meaningful information about a good or service. In turn, this situation would harm the original holder of a trademark that relies on the guarantee of quality that their mark provides in order to sell their products to consumers. But even before the PTO makes its determination, machine learning can also help to optimize the search process from an applicant's perspective, thus providing a role that essentially supplements the PTO's eventual determination by lowering the search costs associated with trademark selection.

While this paper is concerned with the deployment of machine learning in trademark search and registration, it is important to note that a few scholars have analyzed its use in patent applications.[64] In a thoughtful piece about machine learning at the USPTO, Arti Rai discusses the use and implications of its impact in the area of prior art search, noting that it holds significant promise in maximizing efficiency in a world of overburdened patent office administration.[65] While Rai focuses much of her analysis on USPTO reliance on machine learning, her work valuably opens up a larger discussion about the relationship between AI-driven private search engines and the USPTO's own tools.

---

63. WORLD INTELL. PROP. ORG., *supra* note 5, at 107. Outside of the trademark law community, there is a robust conversation ongoing about the future uses of AI for both litigation and transaction-related tasks. *See* John Markoff, *Armies of Expensive Lawyers, Replaced by Cheaper Software*, N.Y. TIMES (Mar. 4, 2011), https://www.nytimes.com/2011/03/05/science/05legal.html; *see also* Timothy J. Carroll & Manny Caixeiro, *Pros and Pitfalls of Artificial Intelligence in IP and the Broader Legal Profession*, LANDSLIDE (Jan. 2019), https://www.dentons.com/en/-/media/fa72a6d5cb304c1194e015eb26123e27.ashx.

Both Rai and Tabrez Ebrahim[66] have noted that AI tools enable patent applicants to design their applications in a way that maximizes their information advantages.[67] Patent applicants have private information about the quality and originality of their patents, and patent examiners must work to uncover this information and make decisions about patentability.[68] Ebrahim valuably explores this idea of information asymmetries between the patent office and the private sector at length. In a model, described as the Spence Model of Information Exchange, he describes a back-and-forth game where the patent applicant and patent office engage in countering signals about the patent's quality.[69] The applicant is always the first mover and will try to maximize the scope of the patent application, and the patent examiner tries to discern whether this scope is reasonable and may try to pare it back.[70] The examiner and patent applicant (or the patent prosecutor) will go back and forth until they settle on an equilibrium.[71] Ebrahim argues that success in this game rests on each party's ability to discover relevant information.[72]

Critically for our study, he also describes how privately supplied AI tools can exacerbate information asymmetries between the patent applicant and the patent office because the patent office does not have the tools to discern between high- and low-quality signals.[73] Thus, the patent office will be in a position where it cannot adequately sift through a market for lemons, thus creating a supply-side issue where the generators of information can more successfully play the information game.[74] More broadly, AI could also displace the need for lawyers, as he explains that:

> [a]rtificial-intelligence technology could displace or reduce the need for attorneys in law firms or in-house legal departments and, in doing so, lessen the job opportunities for law students. The impact of decreasing the role of legal-service professionals with AI technology affects the relationship between clients and lawyers and, as a result, also affects the relationship of the interaction between inventors and the USPTO.[75]

---

64. *See, e.g.*, David Engstrom, Daniel E. Ho, Catherine M. Sharkey & Mariano-Florentino Cuéllar, *Government by Algorithm: Artificial Intelligence in Federal Administrative Agencies* 46–52 (2020), *available at* https://www-cdn.law.stanford.edu/wp-content/uploads/2020/02/ACUS-AI-Report.pdf.

65. Rai, *supra* note 3, at 2619–21; *see generally* Michael D. Frakes & Melissa F. Wasserman, *Irrational Ignorance at the Patent Office*, 72 VAND. L. REV. 975 (2019) (concluding that each patent examiner needs more time to assess a patent application to improve patent quality); U.S. General Accountability Office, *Intellectual Property: Patent Office Should Strengthen Search Capabilities and Better Monitor Examiners' Work*, GAO-16-479 (July 20, 2016), https://www.gao.gov/products/GAO-16-479 (recommending steps to improve the prior art search quality).

66. Ebrahim, *supra* note 3, 104.
67. *Id.* at 1196–1201.
68. *Id.* at 1211–12.
69. *Id.* at 1191.
70. *Id.*
71. *Id.*
72. *Id.* at 1221–23.

We might imagine that similar forces are at play with trademarks. Although trademark approvals, particularly simple word marks, are likely not as complex as patent examinations, there is evidence that AI is transforming this area of IP law as well. The impact of AI on trademark search may be greatest for word marks or composite marks with literal elements, since more data might be available, allowing for greater ease of identifying similarities and differences.[76]

In essence, however, the core search cost problem that Ebrahim and Rai articulate from the perspective of patent applicants and examiners is the same problem that we are exploring from the perspective of trademark applicants. The rise of the private sector in search can have dramatic effects on trademark quality and supply, just like in the patent context. Primarily, the "likelihood of confusion" standard in trademarks is similar to the non-obviousness standard in patents because of the human subjectivity involved in both processes. Each requires an examiner determining whether to grant an application based on their best evaluation of the application, with an eye toward minimizing errors that could result in informational harms to consumers.

Here, we might also note the risk that private vendors' search tools might be more sophisticated than those of the government.[77] Indeed, the emergence

---

   73. *Id.* at 1220.
   74. *See id.* at 1236.
   75. *Id.* at 1231–32.
   76. *Letter from American Bar Association-Intellectual Property Law Section to Secretary of Commerce for Intellectual Property & Director of the United States Patent and Trademark Office*, USPTO (Jan. 9, 2020), https://www.uspto.gov/sites/default/files/documents/ABA-IPL_RFC-84-FR-58141.pdf, at 12 [hereinafter ABA Letter].
   77. She also discusses the risks in relying on private vendors from an explainability/due process perspective, observing that there is at least an appreciable risk that using private search of a private market for trademark search indicates that there may be a market failure regarding trademark registration. Although the USPTO operates its own free search service, there are several private sector alternatives.[78] These private services variously advertise their added value as being powered by AI, machine learning, statistical models, or other sophisticated techniques.[79] Insofar as trademark applicants rely on these private services instead of the USPTO, it suggests that these services provide real value that the government service does not.[80]

Moreover, since the USPTO is not an enforcement agency, and IP rights owners are responsible for protecting their marks, the government may not have the right incentives to have the best AI tools available, and can instead externalize these costs to trademark registrants. This externalization thus creates a market for the sorts of private AI tools in our study, which function to supplement the government's inadequate TESS system. Assuming that the USPTO relies on its own TESS search engine, and that TESS does not work as well as these AI-powered private sector alternatives, the emergence of private search engines suggests that the government's inadequacy may be potentially (indirectly) imposing costs on trademark holders and consumers.

An increase in AI-powered search could plausibly reduce the overall number of applications filed because it would forecast which marks were likely to face a Section 2(d) refusal.[81] Consider: both examiners and applicants want to avoid the monetary and time costs associated with bad applications. A trademark can cost about $250 per class it is registered for,[82] and it takes a substantial amount of time.[83] While the cost of the mark may be trivial for larger companies and brands, the time involved and attorney's fees can be

---

vendors might result in assertions of trade secrecy and more opacity. Rai, *supra* note 3, at 2640–41.

    78. For example, see Corsearch, Markify, Trademarkia, and TrademarkNow. We detail these in a below section.

    79. *See* Nick Potts, *Reviews of the 3 Best Trademark Clearance Search Tools for Trademark Attorneys*, TRADEMARKNOW (Oct. 20, 2016), https://www.trademarknow.com/blog/reviews-of-the-3-best-trademark-search-tools-for-trademark-attorneys.

    80. Part of this extra value-added may come from the fact that the AI technologies underlying trademark search are also used for brand protection. We discuss this further in Part III.

    81. *See* ABA Letter, *supra* note 76, at 11–12.

    82. U.S. Patent & Trademark Office, *Trademark Fee Information*, https://www.uspto.gov/trademark/trademark-fee-information (last visited on Jan. 22, 2021).

    83. *See* U.S. Patent & Trademark Office, *Section 1(b) Timeline: Application Based on Intent to Use your Trademark in Commerce*, https://www.uspto.gov/trademark/trademark-timelines/section-1b-timeline-application-based-intent-use (last visited on Jan. 22, 2021).

substantial.[84] The USPTO provides a useful chart, included as Figure 1, for a 1(b) trademark application—essentially when an applicant files a mark with intent to use it later.[85] At a minimum, from the time an application is filed to when it is approved is about seven months. However, if the USPTO does not immediately approve the mark, it adds at least three months to the process, and as much as an additional eight months if there are multiple rounds of correspondence between the applicant and the USPTO.[86] That additional time could represent lost revenue and other harms stemming from lack of IP protection.

From the USPTO's point of view, AI might provide assistance in achieving greater consistency among Examining Attorneys by helping them reach faster decisions, reducing their workload, and enabling them to identify any inconsistencies in outcomes.[87] It might also aid the detection of fraudulent filings and practices as well, through its evaluation of metadata and closer image comparisons.[88] If AI can be used by applicants to ensure that they do not erroneously file an application that is destined to undergo additional rounds of screening or a final rejection from the USPTO, they can save the time and energy needed to go through the appeals process.

---

84. The examination process involves three steps: first, the mark is classified into a series of design codes; second, examiners search through existing marks, pending applications, and abandoned marks for similarity; and third, issue a determination regarding whether the mark is eligible for registration. *See* Engstrom et al., *supra* note 64, at 47.
85. *Id.*
86. *See Section 1(b) Timeline*, USPTO, https://www.uspto.gov/trademark/trademark-timelines/section-1b-timeline-application-based-intent-use (last visited on Jan. 22, 2021) (Figure 1 below) (showing the timeline for 1(b) applications).
87. *See* ABA Letter, *supra* note 76, at 12.
88. *See id.*

Figure 1: Timeline of Section 1(b) Applications

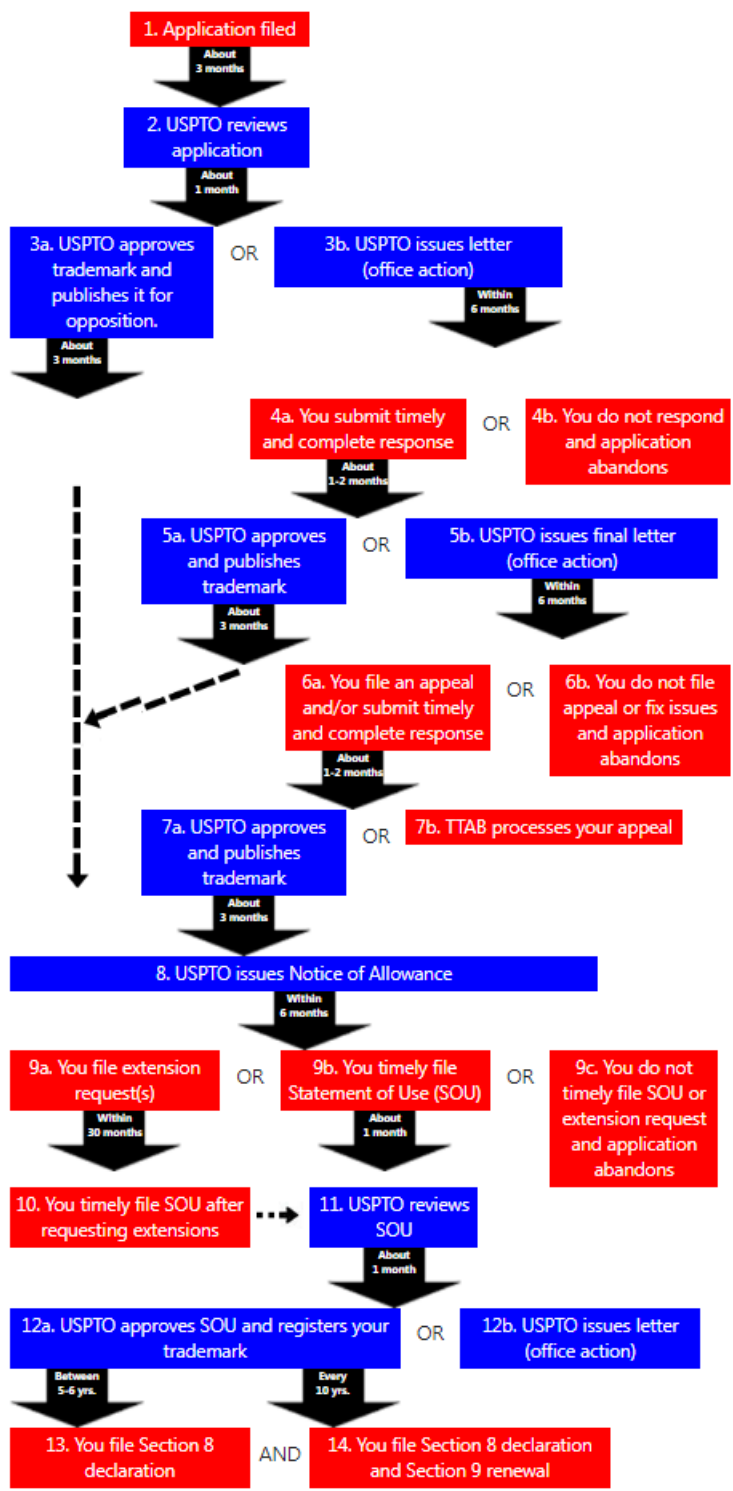

B.     ARTIFICIAL INTELLIGENCE AND THE ADMINISTRATION OF INTELLECTUAL PROPERTY

The prior Section provided a general theoretical basis for the emergence of a private search market for trademarks. A key part of this inquiry involves differentiating between (1) government use of AI-powered techniques to assist with their determinations; (2) AI-powered tools made available by the government to assist private parties in preparing applications for patent, trademark, or copyright protection; and (3) a comparably broader set of AI-driven private tools developed for private parties (rather than the government) in order to supplement state-offered techniques. Below, we focus on the first two categories, in discussing the role of AI in government-led administration of IP and explaining how this paves the way for a private AI-powered market to optimize trademark search and registration. We turn to the third category in our next section.

Last year, WIPO released the first comprehensive global survey of how AI and machine learning can be employed to assist with the governance of IP.[89] Out of WIPO's survey of thirty-five different IP Offices, the report noted that seventeen offices use AI technology in at least one aspect of their work, but that most of these uses appear to be in their infancy.[90] On the patent side, the report points out that AI can be used to automatically analyze the content of patent applications and case files including sorting and allocating them for particular staff, as well as for applying particular classifications.[91] It can also be used for the purposes of searching for prior art and to improve detection of links between citations and applications,[92] and even to assist in the processing of applications.[93] One office in Singapore estimated saving five thousand hours of an examiner's man-hours by relying on AI techniques.[94]

The USPTO is using machine learning in its determinations of patentability and histories of patent prosecution.[95] The USPTO, for example, developed a tool named Sigma, which can search an entire patent document and compare applications with registered patents and pre-grant publications.[96] Rai noted that Sigma enables examiners to attach a particular weight to the most relevant part of the patent application, and then retrieve related documents including related prior art.[97] Another cluster of AI applications are also used to manage IP files' prosecution and formality checks, particularly regarding data support, proofreading, and conversion of files to enhance machine-readability, and also for the purposes of translation and data analysis.[98] According to one study led by David Engstrom, the USPTO is also considering ways to build an AI-driven search platform that would use content-based engines to suggest prior art for an applicant; other plans involve using neural word embeddings to expand prior art searches.[99]

---

89. *See Meeting of Intellectual Property Offices (IPOS) on ICT Strategies and Artificial Intelligence (AI) for IP*, WORLD INTELL. PROP. ORG. (May 23–25, 2018), https://www.wipo.int/edocs/mdocs/mdocs/en/wipo_ip_itai_ge_18/wipo_ip_itai_ge_18_1.pdf.
90. *See id.*
91. *Id.* at 4 (noting developments in Germany, Brazil, and Singapore along these lines).
92. *Id.* at 5–6 (describing patent search systems).
93. *Id.* at 6 (describing Tequmine in Finland for patent classification and prior art search).
94. *Id.* at 12.
95. Isi Caulder & Paul Blizzard, *Canada: Artificial Examiner: The Expanding Use of AI and ML Software at Intellectual Property Offices (IPOs)*, BERESKIN & PARR LLP (July 26, 2018), https://www.bereskinparr.com/doc/artificial-examiner-the-expanding-use-of-ai-and-ml-software-at-intellectual-property-offices-ipos.

On the trademark side, as opposed to patent, the main focus so far has been on the process of search, which historically has been mostly manual.[100] In the context of trademark search, new tools would provide a valuable service by helping potential registrants identify potential conflicts before ever submitting a trademark application, providing the statistical tools necessary to distinguish signal and noise. Colleen Chien, echoing Ebrahim and Rai, has pointed out that the USPTO itself has a difficult time assessing patent quality and frequently grants patents that it probably ought not to.[101] The same might also be said of trademarks, which compels the employment of AI-driven tools in searching for similar marks.

In the context of trademarks, AI has mainly been deployed as an enhancement tool to assess trademark similarity; however, as Dev Gangjee has noted in an excellent study, AI can play a broader, potentially game-changing role.[102] In the government context, it is unlikely that AI will replace human judgment regarding the more complex and subjective tests in trademark law; however, AI still carries the power to streamline administrative tasks relating to registration, opposition, and other procedures, and is likely to only grow in importance.[103]

Traditional search systems employ text-based retrieval technology; today, the technology has improved in order to incorporate phonetic analogies, synonyms, and related permutations of letters in order to compare slightly modified marks as well.[104] Even more, AI has driven significant advances in three additional dimensions of search and comparison: (1) text and conceptual similarity (i.e., assessing text, as well as shared or oppositional meanings of a trademark);[105] (2) visual/image similarity (i.e., assessing image elements of a trademark logo or figurative mark, including content-based image retrieval);[106] and a combination of words and images in order to integrate both in a similarity assessment.[107] Still other approaches rely on a constellation of comparisons—such as automated similarity assessments of image/pixel, text, and content, coupled with a manual comparison—in order to provide a more comprehensive comparison.[108]

---

96. *See* WORLD INTELL. PROP. ORG., *supra* note 89, at 3.
97. Rai, *supra* note 3, at 2634.
98. WORLD INTELL. PROP. ORG., *supra* note 89, at 2, 9–10 (noting developments in Singapore, China, Japan, Morocco, Serbia, and Canada).
99. Engstrom et al., *supra* note 64, at 48.
100. *Id.*
101. *See* Colleen V. Chien, *Comparative Patent Quality*, 50 ARIZ. ST. L.J. 71, 72–74 (2018).
102. Gangjee, *supra* note 6, at 2.
103. Moerland & Freitas, *supra* note 6, at 27.

There is growing evidence of government use of these AI-driven tools as well. Reports indicate that current government uses of AI in the context of trademarks involve image recognition, classification of goods and services, and identifying descriptive terms.[109] According to David Engstrom, the USPTO is also prototyping a deep learning model that uses an unsupervised approach to generate visually similar images from a database.[110] The International Trademark Association has reported that at least five governments have developed trademark image search engines that incorporate AI.[111] The USPTO, for example, has developed a manually coded system of figurative images in order to train its deep learning systems to generate design codes for new trademark image applicants.[112] Other governments rely more extensively on private image search tools for their government registries.[113] For example, IP Australia and the E.U. Intellectual Property Office uses TrademarkVision's Image Recognition (now a part of Clarivate Analytics) to search existing trademark images, employing a technology similar to facial recognition technology, but applied to marks instead.[114] Chile, China, and Japan also rely on private tools.[115] Some offices, such as that of Australia, even offer the public a range of AI-driven tools to assist unregistered applicants.[116] And WIPO's Global Brand Database recently released a free AI-driven image search tool for the public.[117]

---

104. Gangjee, *supra* note 6, at 6 (citing C.J. Fall & C. Giraud-Carrier, *Searching Trademark Databases for Verbal Similarities*, 27(2) WORLD PATENT INFO. 135 (2005)).

105. Gangjee, *supra* note 6, at 6–7 (advances in search technology based on semantic or conceptual similarity focus more on "lexical relations," integrating assessments of synonyms, antonyms, or comparable words in another language) (citing F. Mohd Anuara, R. Setchia & Y-K Lai, *A Conceptual Model of Trademark Retrieval based on Conceptual Similarity*, 22 PROCEDIA COMPUT. SCI. 450, 451 (2013)).

106. Gangjee, *supra* note 6, at 7 (noting that WIPO and the European Intellectual Property Office offer users the ability to upload image-based file formats). Currently, WIPO relies upon a system, the International Classification of the Figurative Elements on Marks, also called the Vienna Classification system. Trademark examiners, in general, manually index and code elements of figurative marks, often in reference to the Vienna Classification system, and then match the Vienna codes of a new application with those already registered. Since not all trademark registries use the system, and it involves some subjectivity, there is the risk of gaps in its application. *See id.* at 7–8 (citing WIPO, Future Development of the Vienna Classification: Questionnaire Results (April 3, 2019)). According to Gangjee, AI-assisted processes of content-based image retrieval have been "welcomed," due to the added value of accuracy in comparison. *Id.*

107. Gangjee, *supra* note 6, at 6–9. As he writes, "[t]he goal is to mimic the assessment of a human examiner who must synthesize visual, aural, and conceptual similarity to arrive at an overall conclusion on whether the marks conflict." *Id.*

108. *Id.* at 10 (citing Mosseri I., Rusanovsky M. & Oren G., *TradeMarker – Artificial Intelligence Based Trademarks Similarity Search Engine*, in COMMUNICATIONS IN COMPUTER AND INFORMATION SCIENCE (vol. 1034, 2019), https://doi.org/10.1007/978-3-030-23525-3_13); Moerland & Freitas, *supra* note 6, at 2 (noting that only a few trade mark offices apply AI tools).

109. Moerland & Freitas, *supra* note 6, at 15; *see also* Engstrom et al., *supra* note 64, at 49 (describing the use of a deep learning image classifier and other prototypes).

The wide range of emerging tools may lead some to suggest that AI might even have the effect of shrinking the potential role of the trademark lawyer. Since automated technologies can play a wider role in brand clearance and brand protection, it would enable service providers to work directly with trademark owners themselves. Echoing this view, others have observed that AI's added efficiency has the potential to replace paralegals or junior lawyers, perhaps when it comes to search and registration.[118] However, more complex situations still call for human intervention. One WIPO survey respondent from Norway was careful to note that in comparing AI and non-AI results, while the most "similar" trademarks often had the same results, there were very large differences found between AI and non-AI results in addressing lower degrees of similarity.[119] This suggests that a mix of human and non-human intervention and greater amounts of data would improve the outcome.[120]

---

110. Engstrom et al., *supra* note 64, at 49–50 (also describing future ways to deploy AI in image/text classification).

111. *See INTA Comments in Response to Request for Comments on Intellectual Property Protection for Artificial Intelligence Innovation*, USPTO 1, https://www.uspto.gov/sites/default/files/documents/International%20Trademark%20Association%20(IN_RFC-84-FR-58141.pdf (noting that out of 9 respondents to its survey, five IP offices are using AI-driven tools in trademark image search systems).

112. Gangjee, *supra* note 6, at 9 (citing U.S. Patent Trademark Office, Emerging Technologies in USPTO Business Solutions (May 25, 2018), https://www.wipo.int/edocs/mdocs/globalinfra/en/wipo_ip_itai_ge_18/wipo_ip_itai_ge_18_p5.pdf).

113. Gangjee, *supra* note 6, at 9.

114. *Id.*

115. WORLD INTELL. PROP. ORG., *supra* note 89, at 7–8 (describing developments).

116. *Id.*

117. *See Global Brand Database*, WORLD INTELL. PROP. ORG., https://www3.wipo.int/branddb/en/ (last visited July 8, 2020).

118. *See How AI Impacts Trademarks*, TRADEMARK TIMES 1 (2018), https://www.managingip.com/pdfsmip/01-TrademarkTimes18Seattle.pdf.

Beyond image search, offices reported relying on AI techniques for the purposes of trademark examination as well. Australia uses a Smart Assessment Toolkit that relies on natural language processing and internal software to detect substantially similar trademarks, and the office in Singapore uses machine learning techniques to measure and suggest parameters to measure trademark distinctiveness.[121] Of course, like other areas of AI applications, there are significant risks associated with automated decision-making, some of which stem from the legal and cultural risks associated with lack of transparency, unrepresentative training data, or difficulty in explainability, which we address more below.[122] Particularly in the context of trademark law, which relies on subjective, context-dependent assessments, AI-driven technologies may be less useful in terms of evaluating distinctiveness, likelihood of confusion, and other variables that require a nuanced evaluation.[123]

Of course, one additional consideration for the success of AI in trademark law involves the need for accurate, structured, multi-jurisdictional and comprehensive data.[124] Towards this end, scholars Anke Moerland and Conrado Freitas have distinguished between two different types of data: legal data, that involves decisions, oppositions and invalidity proceedings, and case law from various jurisdictions, in order to improve the accuracy of legal predictions; and market-based data, which includes information about consumer preferences, product variations, goodwill, product reputation, distinctiveness, etc.[125] As they note, privacy and data protection laws can impede the collection of such data, making both types of data difficult to compile accurately and comprehensively (let alone across jurisdictions), thereby posing a challenge to the efficacy of AI-driven judgments in the global trademark ecosystem.

---

119. Moerland & Freitas, *supra* note 6, at 15.
120. WORLD INTELL. PROP. ORG., *supra* note 89, at 8.
121. *Id.*
122. *See id.* at 12 (noting that Australia has developed an Automated Decision-Making Governance Framework and Policy); *see also* Engstrom et al., *supra* note 64, at 50–51 (noting explainability concerns, among others, in deploying AI at the USPTO).
123. Moerland & Freitas, *supra* note 6, at 16.
124. *Id.*
125. *Id.*

C. ARTIFICIAL INTELLIGENCE IN PRIVATE TRADEMARK SEARCH AND REGISTRATION

As studies have postulated, AI carries the potential to revolutionize advertising, particularly in terms of consumer recommendations, targeted advertising, market forecasting, and speech and text recognition.[126] However, AI-related issues have been largely underexamined regarding trademarks, specifically, especially where legal doctrine is concerned.[127] Just recently in late 2019, the USPTO solicited public comments about a range of issues involving AI and IP, including issues surrounding patents, authorship and copyrightability, trademark registrability, and datasets, amongst others.[128]

While the vast majority of comments received focused on copyright and data-related issues, several consistent themes emerged regarding trademark protection. As Dev Gangjee has explained, the effect of AI on trademark registration will be more subtle than its impact on copyright or patent law, which has largely been driven by a threshold question of whether autonomous agents can be considered authors or inventors and whether the resulting work product is protectable.[129] Very few comments focused on the related question of AI-created marks. At least one commentator concluded that the possibility of AI-created marks existed but emphasized that only live humans should be able to file for registration.[130]

---

126. *See The Future of Trademark Service Providers*, WORLD TRADEMARK REV., https://www.worldtrademarkreview.com/reports/the-future-of-trademark-service-providers (last visited Jan. 23, 2021) (portions on file with author) [hereinafter "TM Report"]; *see also* Interactive Advertising Bureau, *Artificial Intelligence in marketing Report*, IAB (Dec. 9, 2019), https://www.iab.com/insights/iab-artificial-intelligence-in-marketing/; Lee Curtis & Rachel Platts, *AI is Coming and It Will Change Trade Mark Law*, MANAGINGIP (Dec. 8, 2017), https://www.hgf.com/media/1173564/09-13-AI.PDF (focusing mostly on trademark law and its effect on retail, also noting how the law must adapt to AI); Lee Curtis & Rachel Platts, *Trademark Law Playing Catch-up with Artificial Intelligence?*, WIPO MAG. (June 2020), https://www.wipo.int/wipo_magazine_digital/en/2020/article_0001.html (same); Yashvardhan Rana, *Artificial Intelligence and Trademark Law in the Digital Age*, INTERNATIONAL JURIST (July 29, 2020), https://www.nationaljurist.com/international-jurist/artificial-intelligence-and-trademark-law-digital-age#:~:text=Such%20products%20also%20enable%20a,in%20turn%20saving%20lawyers'%20time (discussing the potential effect of AI on trademark law). Recommendation systems might also arguably spark trademark liability claims if they offer competing products to a consumer, stemming from theories of initial interest confusion. Here, the jurisprudence on keyword searches can be instructive, as well as recent case law questioning the reach of initial interest confusion, suggesting that such theories of liability are unlikely to succeed in court. Gangjee, *supra* note 6, at 1–2. *See, e.g., Multi Time Mach., Inc. v. Amazon.com, Inc.*, 804 F.3d 930 (9th Cir. 2015) (noting that clear labels by Amazon in making recommendations precluded a theory of liability); *Rescuecom Corp. v. Google Inc.*, 562 F.3d 123 (2d Cir. 2009) (Google's use of the Rescuecom trademark was a use in commerce); *Rosetta Stone v. Google*, 676 F.3d 144 (4th Cir. 2012) (overturning a grant of summary judgment for Google).

127. *See* Gangjee, *supra* note 6.

128. *Request for Comments on Intellectual Property Protection for Artificial Intelligence Innovation*, USPTO (Oct. 30, 2019), https://www.federalregister.gov/documents/2019/10/30/2019-23638/request-for-comments-on-intellectual-property-protection-for-artificial-intelligence-innovation.

The vast majority of trademark-related comments from organizations concluded that the use of AI would improve and streamline the trademark search and registration process, noting that "[d]ecisions to proceed or not to proceed with filing a U.S. trademark application for a particular mark may be made more quickly and may be better informed if driven by a more objective risk assessment."[131] However, the increase in accuracy, at least one commentator noted, would also raise the bar for successful trademark applications, making them potentially harder to obtain but improving the overall quality of trademarks nevertheless.[132]

A second theme was that AI could have a transformative effect on the detection of trademark infringement with its rapid search and comparison technology, aiding the USPTO in determining fraudulent applications.[133] At the same time, the American Bar Association (ABA) also noted the risk that AI tools and software could be used in the opposite way—to infringe the rights of other trademark owners—thus opening up questions of machine volition and liability.[134] At least one other commentator expressed a similar view, warning that while AI could be used to better detect infringement and protect trademarks, the very same technology could also be used to violate trademark rights as well.[135] The commentary, from a center focused on anti-counterfeiting, warned that AI could be used to detect gaps in trademark protection and deceive consumers with strategically driven recommendations.[136] "At what level of prediction is there a duty to inform consumers, or b[r]and owners, about a potentially suspicious product?," the commentary asked, noting a potentially increased risk of inaccuracy from AI-driven counterfeit detection.[137] Here, if an AI tool makes an infringing recommendation, consumer harms might stem not from initial interest or point-of-sale confusion, but rather from the harm of post-sale confusion.[138]

---

129. Gangjee, *supra* note 6, at 1.
130. *See Commentary from A-CAPP*, USPTO 1 (Dec. 16, 2019), https://www.uspto.gov/sites/default/files/documents/Jeffrey-Rojek_RFC-84-FR-58141.pdf (noting that "the creation of the trademark itself should not be allowed by AI, emphasizing role for humans in registration").
131. ABA Letter, *supra* note 76, at 5; *see also Letter from Computer & Communications Industry Association and Internet Association to Secretary of Commerce for Intellectual Property & Director of the United States Patent and Trademark Office*, USPTO 10 (2020) (noting searches would be faster and more efficient) [hereinafter Computer & Communications Industry Letter]; Trevor Little, *Lower risk applications, increased refusals and a boost for infringers: the potential impact of AI on trademarks*, WORLD TRADEMARK REV. (Mar. 23, 2020), https://www.worldtrademarkreview.com/anti-counterfeiting/lower-risk-applications-increased-refusals-and-boost-infringers-the.
132. *See generally* Letter from Obeebo, Inc. to Secretary of Commerce for Intellectual Property & Director of the United States Patent and Trademark Office, USPTO, https://www.uspto.gov/sites/default/files/documents/Obeebo-Inc_RFC-84-FR-58141.pdf (noting that AI will raise the bar for distinctiveness, but ultimately improve trademark quality).
133. *See* ABA Letter, *supra* note 76, at 12 (noting that AI could aid a pixel-by-pixel comparison); *Comments from the App Association*, USPTO 5 (date goes here), https://www.uspto.gov/initiatives/artificial-intelligence/notices-artificial-intelligence-non-patent-related (noting that AI tools are used to detect infringement).
134. ABA Letter, *supra* note 76, at 13.

A third theme involved the consistent idea that the law did not need reforming due to the advent of AI, although many expressed a desire to avoid weakening trademark protection as a result of AI.[139] One representative view, along similar lines, expressed by the ABA and several others, involved the conclusion that AI could serve as "an appropriate supplement, but not a substitute for the human judgment of [counsel]."[140] Similarly, another set of commentators observed that using AI to supplement (rather than supplant) human judgment would avoid the risk that complete reliance on AI might produce an incorrect conclusion.[141] At least one study echoed this view by noting that the subjectivity and complexity of trademark law's doctrinal tests would be difficult to replicate with an AI-driven system, since they are presently unable to reflect the nuances of these tests.[142]

---

135. *See Comments from the Center for Anti-Counterfeiting and Product Protection*, USPTO (Dec. 16, 2019), https://www.uspto.gov/sites/default/files/documents/Jeffrey-Rojek_RFC-84-FR-58141.pdf.

136. *Id.* at 2–3.

137. *Id.* at 3.

138. *See* Trevor Little, *Lower Risk Applications, Increased Refusals and a Boost for Infringers: The Potential Impact of AI on Trademarks*, WORLD TRADEMARK REV. 2 (Mar. 23, 2020), https://www.worldtrademarkreview.com/anti-counterfeiting/lower-risk-applications-increased-refusals-and-boost-infringers-the (quoting commentary from the American Intellectual Property Law Association).

139. *See Comments from the App Association*, *supra* note 133, at 5 (noting a desire to avoid weakening trademark law); *see also* Computer & Communications Industry Letter, *supra* note 131, at 10 (noting no impact of AI on trademark law, and no need to change the law at this time).

140. ABA Letter, *supra* note 76, at 12 (noting that AI should not be used as a substitute for subjective judgment); *see also Letter from IBM Corporation to Secretary of Commerce for Intellectual Property & Director of the United States Patent and Trademark Office*, USPTO 5 (Jan. 19, 2019), https://www.uspto.gov/initiatives/artificial-intelligence/notices-artificial-intelligence-non-patent-related (noting that a trademark examiner will still be required to assess the evidence collected in the examination and registration process).

141. *Letter from Japan Intellectual Property Association to Secretary of Commerce for Intellectual Property & Director of the United States Patent and Trademark Office*, USPTO 2 (Jan. 8, 2020), https://www.uspto.gov/initiatives/artificial-intelligence/notices-artificial-intelligence-non-patent-related; *see also* Intellectual Property Owner's Association 6, *available at* https://www.uspto.gov/initiatives/artificial-intelligence/notices-artificial-intelligence-non-patent-related.

Yet most commentary noted, as applied specifically to trademarks, AI carries perhaps the strongest potential in areas of private search and registration.[143] More recent tightening of corporate budgets, coupled with improvements to AI technology, have streamlined the potential for AI to have a transformative effect on the process of trademark registration and litigation.[144] Here, AI-powered search takes a form that is much more predictive in nature, since it is primarily concerned with giving a potential registrant information about whether a preexisting registration will cause their application to be rejected. This type of search can range in complexity. At its most basic, a search engine might check to see if an application exactly matches an existing registration. More complex implementations might use AI to determine the likelihood that the USPTO would reject an application by modeling their own decision-making process. Other techniques might be most advantageous when they can be used to automate tasks like trademark search and watch results.[145] Since AI provides great improvements in terms of speed and accuracy, it can dramatically assist brands who aim to be the first to reach the market.[146]

While a comprehensive view of all of the implications of AI for trademark law is beyond the scope of this article, it bears mentioning that we can envision at least five different ways in which AI-related technologies can radically alter our existing legal systems, and drive the processes of search and registration to be much more proactive in terms of identifying variables that can prove determinative later on.[147] Some examples involve the following:

---

142. Moerland & Freitas, *supra* note 6, at 2.
143. *See* TM Report, *supra* note 126, at 1 (page number corresponds to excerpts on file with author).
144. *See id.* at 3 (page number corresponds to excerpts on file with author).
145. *See* Rob Davey, *Artificial Intelligence: A Meeting of Minds*, WORLD TRADEMARK REV. (Nov. 1, 2017), https://www.worldtrademarkreview.com/portfolio-management/artificial-intelligence-meeting-minds.
146. In particular, models that draw on fuzzy logic are particularly well suited for knowledge that contains elements of vagueness, like knowledge based on natural language. Anna Ronkainen describes how type-2 fuzzy logic systems are particularly appropriate for representations of second-order vagueness, especially in situations, like trademarks, where there may be a "vagueness of a concept and [an] uncertainty associated with its application." Anna Ronkainen, *MOSONG, a Fuzzy Logic Model of Trade Mark Similarity*, in PROCEEDINGS OF THE WORKSHOP ON MODELING LEGAL CASES AND LEGAL RULES 23–25 (Adam Z. Wyner ed., 2010). In simple terms, Ronkainen writes, "traditional fuzzy logic allows us to say that John is 0.9 TALL (whatever that means), whereas with type-2 fuzzy logic we can also say that John is between 0.85 and 0.95 (0.90 +/- .05 TALL), in which the uncertainty or margin of error may stem from any source, anything from potential measurement errors to intrinsic design factors within the model." *Id.*

### 1. *Search, Identification, and Suggestion*

AI carries the potential to help trademark owners search and identify potential trademarks for registration by employing AI to study a wide range of variables relevant to the search process including sight, sound, visual cues, classification of goods/services, and other trademark attributes like descriptiveness. But this can also integrate other external considerations in its analysis, like identifying geographic areas of potential growth, obstacles for trademark goodwill, other similar trademarks, or by noting attributes of other firms within the trademark ecosystem.

The same observation can easily be made for the role that AI and machine learning techniques play in the process of trademark selection.[148] AI can direct the trademark firm applicants to various options that are curated for them, drawing from a vast expanse of market-based data on consumer preferences, brand equity, common law variations, linguistic sophistication, natural language associations, and the like. Search and registration can also be improved using AI techniques, where machine learning can be relied upon to identify semantically similar marks.

### 2. *Registration and Clearance*

AI carries the potential to revolutionize the process of registration, both in terms of automating the processes of registration and in terms of identifying particular areas where there may be conflicting registrations, and even drafting initial registrations or filings and general portfolio management.[149] An expert notes,"[B]rand owners will be able to clear a campaign in weeks or even days, which is essential given how quickly products and services are developed and expand."[150] Another expert adds, "Naming decisions will happen in real time."[151] As these comments suggest, not only can tools "clear" certain proposed marks for registration, but they can also register marks with automated tools.

---

147. *See* TM Report, *supra* note 126, at 4 (page number corresponds to excerpts on file with author).
148. *See* Moerland & Freitas, *supra* note 6, at 4 (describing how machine learning operates in the trademark context).
149. *See* TM Report, *supra* note 126, at 3 ("Areas where AI will dominate include searching and clearance, prosecution (at least for simple marks), renewals and possibly even oppositions.") (page number corresponds to excerpts on file with author).
150. *See id.* at 4 (page number corresponds to excerpts on file with author).
151. *Id.* (page number corresponds to excerpts on file with author).

### 3. *Comparison and Determining Substantial Similarity*

AI can alter the processes of investigating substantial similarity by relying on deep learning and fuzzy logic techniques to evaluate comparisons of trademarks and product attributes. It can investigate multiple types of similarity—visual, semantic, and image—in seconds.[152] By using neural network technologies, entities can process large amounts of data in order to determine semantic equivalence, providing insights into substantial similarity and trademark relatedness.[153] As Anna Ronkainen further explains:

> Trademark similarity search . . . requires searching for dissimilar images as opposed to the more common approach of searching similar (or identical) images. In the latter, as long as the amount of similar images is sufficient, one could try to train a neural network-based model to catch similarities between images. For example, in order to teach the machine to differentiate between cats and dogs we should supply it with many images of cats and dogs. Unfortunately, in a trademarks database, this is obviously not the case. Moreover, catching differences between trademarks is far more complex since it is much harder to find pairs of similar trademarks, and on top of that, there is no formal definition of similar trademarks, as trademarks are considered to be similar only if they are *deceptively* similar.[154]

As she notes, while there are some difficulties with training machines to capture these complexities, it is reasonable to consider that techniques will continue to improve in time, thereby assisting with the determination of substantial similarity.[155] Others have expressed similar concerns, noting that determining trademark distinctiveness, the relevant public, the proper classification of goods and services, among other elements, are so subjective that they pose challenges to the development of AI in trademark law.[156]

---

152. Visual similarity involves the question of whether two trademarks are visually similar; semantic similarity involves whether the trademarks contain the same meaning and semantic content; and text similarity involves whether the actual text of the trademark is similar. Idan Mosseri et al., *How AI will Revolutionise Trademark Searches*, WORLD TRADEMARK REV. (July 2, 2019), https://www.worldtrademarkreview.com/ip-offices/how-ai-will-revolutionise-trademark-searches.

153. *See generally* TM Report, *supra* note 126 (excerpts on file with author).

154. *See* Ronkainen, *supra* note 146, at 23–25 (discussing the difficulties in training an AI program to catch differences between trademarks).

155. *Id.*

156. Moerland & Freitas, *supra* note 6, at 20–23.

#### 4. Prediction and Risk Assessment

As with each of the other areas, the real payoff of AI lies in its ability to predict the outcomes of various trademark-related decisions—such as the litigation risk involved in proceeding with a particular trademark or product—and the market implications of making certain choices.[157] Risk assessments are very useful; as Gangjee notes, "[w]hile human expertise continues to assess the conflicts results lists generated by algorithms, for risk-averse commercial clients it is extremely tempting to be guided by clearly defined percentages of similarity."[158] Indeed, predictive analytics can prove to be transformative in helping businesses both create and sustain a strong presence in the marketplace, predicting the outcome of filing suit, sending a cease-and-desist, articulating various claims, or deciding whether and for how much to settle. And this is just the tip of the iceberg. Imagine every aspect of a trademark claim—its probable outcome automated, calculated, predicted and ready for real-time decision-making.

Nevertheless, despite the improvements AI will provide regarding trademark registration and litigation, it is important to note that experts continue to emphasize the importance of human oversight and participation, particularly in terms of using human judgement in complex cognitive tasks, especially in the context of trademark doctrines which are highly context-specific. This is especially true in more complex cases of multi-word or slogan marks, where humans are likely to be the best at determining areas of particular strength.[159]

#### 5. Brand Management

Finally, nearly every private trademark search engine company in our study offers brand protection services in addition to their trademark search services in some capacity.[160] These brand protection services generally include some combination of active monitoring of U.S. and global databases, and sometimes

---

157. *See* TM Report, *supra* note 126, at 4 (page number corresponds to excerpts on file with author).
158. *See* Gangjee, *supra* note 6, at 13.
159. *See generally* TM Report, *supra* note 126 (excerpts on file with author). One example of this, experts suggest, is having a team of humans who can physically review and correct the data from national trademark registries to ensure that proprietary trademark databases have correct examples, deleting, for example, cases where the word mark does not match the image (errors which are easy for automated systems to overlook). *See generally id.*
160. *See, e.g.*, *Quickly respond to potentially infringing trademark applications with a powerful suite of watch solutions*, COMPUMARK, https://www.compumark.com/solutions/trademark-watching/watching (discussing CompuMark's trademark watching services).

tools for pursuing legal enforcement of trademark rights against potential infringers.[161]

At a later phase of search, current trademark holders might engage in a proactive process of brand management, vigilantly searching for newly registered marks that may threaten to dilute the strength of the older trademark holder's mark. Because of the huge search costs in finding potentially conflicting trademarks, trademark owners could face a daunting proposition in attempting to enforce their trademark rights themselves. This is essentially the same problem that confronts potential registrants, who must filter out the noise and recover actual conflicts, as we have previously asserted in this paper.

Here, again, as we have suggested, AI and machine learning techniques can offer mark owners a substantial advantage in brand management and enforcement. We have strong theories about why trademarks are valuable for owners and consumers; they reduce the friction created by information asymmetries and thus facilitate useful transactions.[162] Brand management is important because trademark owners need to maintain the strength of their marks in order to reduce information asymmetries.[163] Moreover, brand protection is a critical service because the USPTO explicitly says that it is not responsible for trademark enforcement; it explicitly places this burden on trademark holders.[164]

Regardless of the reason, the additional benefit that these firms provide to their clients fits into the broader story of how the private sector is able to utilize AI in a way that gets ahead of government resources, supplementing when needed. This is discussed further below.

## III. A COMPARATIVE ASSESSMENT OF THE PRIVATE SECTOR IN TRADEMARK SEARCH

One of the reasons we decided to write this Article is related to another overall observation: aside from a few prominent, recent pieces,[165] there is not a great deal of empirical research on trademark ecosystems, especially compared to other areas of IP. Moreover, while trademark law as a field of study has been thoroughly theorized, there is little to no systematic evidence that compares the various private vendors in the process of trademark search and registration. One relatively recent study identified fewer than seventy articles involving empirical analysis of trademarks.[166] While some areas involved studies of the relationship between trademarks, innovation, and firm performance, the relevant law review literature is still somewhat thin.[167] Other empirical pieces in trademark law have focused on questions of scarcity,[168] the extent of trademark dilution,[169] or the relationship between trademarks and innovation.[170]

---

161. *See infra* Section III.B.2 (full descriptions of each search engine).
162. *See generally* Wessells, *supra* note 16.
163. *See generally id.*
164. U.S. PATENT AND TRADEMARK OFFICE, PROTECTING YOUR TRADEMARK: ENHANCING YOUR RIGHTS THROUGH FEDERAL REGISTRATION 3 (2019) ("You, as the mark owner, are solely responsible for enforcement [of your trademark].").
165. Rai, *supra* note 3; Ronkainen, *supra* note 146.

The problem of firm search costs in trademark search, therefore, lies at the periphery of these various literatures, but there is very little concrete evaluation of the issue. For example, we could find only one other study that considers how different vendors use machine learning techniques in search and registration (and this one focused mostly on government tools).[171] Computer science literature implicitly recognizes that firms face search costs in finding potential conflicts and attempts to optimize methods that reduce these costs,[172] but it does not delve into the economic consequences of deploying these methods. Similarly (and conversely), economics literature implies that the USPTO plays an important gatekeeping function in ensuring adequate search quality (i.e., that potentially damaging marks are not registered),[173] but has never addressed the question of how private vendors have emerged to respond to the USPTO's own search limitations.

Like social science literature, computer science literature is largely theoretical. Authors are primarily concerned with optimizing search algorithms and engines, rather than evaluating current implementations. There do not seem to be meta-studies that comprehensively evaluate either the visual-based search engines or text-based ones, which suggests that there are avenues for future research. The computer science literature is more directly concerned with the efficacy of search algorithms, focusing on more complex problems than standard text-based retrieval. In general, the major problem currently being tackled in the computer science literature is improving trademark retrieval based on visual similarity, instead of spelling or phonetic similarity. Various papers explore different similarity metrics that determine whether proposed design marks are similar to current marks.[174] One 1987 study looked at a system that could return trademark searches based on phonetic similarity, and this application is closer to the technology deployed by the search engines in our study.[175]

---

166. *See generally* Shukhrat Nasirov, *The Use of Trademarks in Empirical Research: Towards an Integrated Framework* (Nov. 20, 2018) (unpublished manuscript), *available at* https://papers.ssrn.com/sol3/papers.cfm?abstract_id=3296064.

167. *See id.*

168. *See, e.g.*, Beebe & Fromer, *supra* note 56, at 947.

169. Paul J. Heald & Robert Brauneis, *The Myth of Buick Aspirin: An Empirical Study of Trademark Dilution by Product and Trade Names*, 32 CARDOZO L. REV. 2533, 2574–75 (2011).

170. Nasirov, *supra* note 166.

171. *See generally* Moerland & Freitas, *supra* note 6.

172. *See* Fatahiyah Mohd Anuar, Rossitza Setchi & Yu-Kun Lai, *A Conceptual Model of Trademark Retrieval Based on Conceptual Similarity*, 22 PROCEDIA COMPUT. SCI. 450, 451 (2013) ("[I]n the Internet age, it is even more important to have efficient mechanisms for protecting trademarks and tools for detecting possible cases of infringement" to motivate the importance of developing their trademark conceptual similarity model.).

173. *See generally* Landes & Posner, *supra* note 42 (framing the economics of trademark law as being grounded in the economics of property and tort law). They argue that trademark creates a property right, and trademark litigation is a branch of tort law. *Id.* Since the USPTO grants the mark, it effectively is responsible for determining who gets a property right.

In terms of literature that directly looks at the applications of text-based trademark retrieval, there are very few. Anke Moerland and Conrado Freitas conducted a study that combined qualitative methods with small-n search tests on the United Nations, European Union's, Australia's, and Singapore's public-facing trademark search engines. Moerland and Freitas note that each of these offices currently uses AI methods to power their search algorithms, whereas the USPTO is still developing and testing AI tools to identify grounds for refusals to register trademark applications.[176] While the study was highly illuminating, it was driven more towards examining government use of AI-related tools (as opposed to studying the private market for assessing trademark search). The study also assessed the functionality of these tools in identifying and comparing visual and conceptual similarity, descriptiveness, morality, and classifications of goods between marks.[177]

In terms of other studies, one 1999 paper examined the potential future of patent and trademark librarians in a time when databases were becoming more common.[178] The author ultimately concluded that librarians would still have a place in helping users navigate these databases, in part because of the huge volume of applications.[179] However, there are no retrospective studies that indicate how this prediction bore out, or how the growth of patent and trademark databases have altered applications or research.

---

  174. *See generally* Anuar, *supra* note 172; Anil K. Jain & Aditya Vailaya, *Shape-Based Retrieval: A Case Study with Trademark Image Databases*, 31 PATTERN RECOGNITION 1369 (1998); Gianluigi Ciocca & Raimondo Schettini, *Similarity Retrieval of Trademark Images*, *in* PROCEEDINGS 10TH INTERNATIONAL CONFERENCE ON IMAGE ANALYSIS AND PROCESSING 915 (Bob Werner ed., 1999).
  175. J. Howard Bryant, *USPTO's Automated Trademark Search System*, 9 WORLD PAT. INFO. 5 (1987).
  176. Moerland & Freitas, *supra* note 6, at 6 (discussing methodology). The qualitative semi-structured survey was sent to fourteen stakeholders, including TrademarkNow and the USPTO, which we include in our assessment. The search engine tests involved searching marks related to Apple Inc., using both its logo and the word "apple" to see if each engine flagged potential issues with the search. Specifically, they measure whether searching "apple" raises conceptually similar marks across all trademark classes and within specific classes like "fruits" and "software/hardware." As we detail below, we conduct similar tests on private sector trademark search engines using over a hundred search terms.
  177. *Id.* at 7–8.

In one related study, Lisa Larrimote Ouellette directly tackles the question of the PTO using search engines in trademark applications. Her main argument is that Google is an underexplored tool in assessing the distinctiveness of a trademark, where distinctiveness is "the extent to which consumers view a mark as identifying a particular source."[180] She argues that Google, with its complex algorithm and public results, provides an easy way for the PTO to assess distinctiveness in cases of infringement. To prove her argument, she conducts an empirical experiment where she used trademarks that were disputed for trademark infringement and searched them through Google.

The basic test for whether a trademark was distinctive in this framework came down to whether it was findable in Google. If a mark was distinctive or commercially popular, then it would dominate the top ad results. If the mark was likely to be confused with another mark, there would be overlapping results between searches for those marks. Essentially, she argues, valuably, that Google can take a lot of the guesswork out of determining whether consumers would be able to discern one mark from another, and that this potential role has been underexplored in infringement cases.

Ouellette's insightful study foregrounds the role of private companies in facilitating search and comparison and points out the potentially powerful (and troubling) role of "algorithmic authority" in trademark law.[181] This study differs from ours primarily in that we are focused on search engines that specialize in trademark search, prior to registration, and we approach the problem from the perspective of a trademark applicant, as opposed to a traditional consumer. Ouellette's solution is mainly relevant for courts deciding a trademark infringement case, whereas we are examining trademark applications well before they would get to that stage of the legal process.[182]

---

178. Julia Crawford, *Obsolescence or Opportunity? Patent & Trademark Librarians in the Internet Age*, 21 WORLD PAT. INFO. 267 (1999).

179. *Id.*

180. Lisa Larrimore Ouellette, *The Google Shortcut to Trademark Law*, 102 CALIF. L. REV. 351 (2014).

181. *See id.* at 368 (citing Clay Shirky's observations) (citation omitted).

182. Moreover, Google differs from trademark search engines in that Google's PageRank algorithm relies on a calculation of how different webpages point to each other. *See How Google's Algorithm Rules the World*, WIRED (Feb. 22, 2010), https://www.wired.com/2010/02

Trademark search engines differ in that they are more akin to querying a database, modeling good results, and returning those results to the user at a much earlier stage involving trademark selection and application. Nevertheless, her observations about the potential role of algorithms in assisting legal determinations are salient to our study, since many private search engines raise similar questions about efficacy, impact, and accuracy.

Other than the ones mentioned above, we could find only two other papers that conducted empirical tests on actual trademark search engines. Anna Ronkainen conducted a study of thirty thousand trademarks conflicts on the TrademarkNow platform.[183] She specifically models a trademark similarity algorithm developed by Onomatics, a Finland-based legal technology firm.[184] The Onomatics algorithm was used to power TrademarkNow's Namecheck product, and she argues it is especially good at incorporating the role of goods and services in its similarity calculation.[185] Her basic results showed that the algorithmic approach recovered marks with precision of about 80% and recall of about 94.9%.[186] A paper entitled "Trademark Search Tools" put forward by ipPerformance in 2011 is the only one that directly looks at leading trademark search vendors and does an apples-to-apples comparison of them.[187]

A.  TRADEMARK SEARCH AND REGISTRATION PROCEDURES

As discussed earlier, potential trademark registrants file their trademarks by filing an application with the USPTO. The USPTO advises that registrants should first determine whether a trademark is the appropriate protection (instead of a patent or copyright), and then details several steps for registration.[188] In particular, it says that registrants should select their mark, choose a format, identify whether it is a good or service mark, search for potential conflicts, and choose a filing basis.[189]

---

/ff_google_algorithm/ [https://web.archive.org/web/20140412235725/http://www.wired.com/2010/02/ff_google_algorithm/].

183. Anna Ronkainen, *Intelligent Trademark Analysis: Experiments in Large-Scale Evaluation of Real-World Legal AI*, in PROCEEDINGS OF THE 14TH INTERNATIONAL CONFERENCE ON ARTIFICIAL INTELLIGENCE AND LAW 227 (Ass'n for Computing Mach. ed., 2013).

184. *Id.*

185. The trademark similarity algorithm is derived from the MOSONG prototype, which is a model of vagueness and uncertainty in legal text. *Id.* at 2.

186. *Id.*; *see also infra* Section III.B.7 (formal descriptions of terms used here).

187. IPPERFORMANCE GRP., TRADEMARK SEARCH TOOLS: ANALYSIS PAPER 7 (2011), https://www.markify.com/pdf/Trademark_Search_Tools_Analysis_Paper-P2a.pdf.

188. *See Trademark Basics*, USPTO, https://www.uspto.gov/trademarks-getting-started/trademark-basics.

189. *Id.* ("Other initial considerations").

Once these steps are complete, the registrant fills out an application, specifying both the mark and the class of goods upon which it will appear, and then monitors its status for USPTO approval. Crucially, each mark must be categorized as either a good or service, and the applicant must select the number of classes.[190] Each class costs an additional $225–275 depending on the specific applicable fee schedule.[191] It may be necessary to communicate with a USPTO examining attorney to talk through any potential issues or objections before getting an official approval or denial. If approved, the registrant is still responsible for enforcing the trademark.[192]

As we well know, in the conventional case, trademark registrants will want to avoid the costs associated with filing a rejected trademark application. To avoid incurring these costs, they turn to trademark search engines to identify potential conflicts in advance and to make appropriate changes prior to filing a trademark application. The first step is generally to check the USPTO's TESS.[193] However, while TESS can return existing trademarks that are similar to the search term, as we have suggested, it is not totally effective. The USPTO itself recommends consulting an attorney before filing an application as it cannot guarantee that its results will be exhaustive.[194]

In the typical use case for a trademark search engine, a potential registrant, or their attorney, searches a potential mark and then sorts through the returned results. Firms may employ attorneys to conduct a trademark search, and, consequently, attorneys turn to trademark search engines to assist with this process. Attorneys need to be exceptionally careful when advising their clients,

---

190. *See* Engstrom et al, *supra* note 64, at 46–47 (description of the trademark process before the USPTO).

191. For more details on the trademark application form, see *Trademark Initial Application Form*, USPTO, https://www.uspto.gov/trademarks-application-process/filing-online/initial-application-forms#Chart%20Application%20requirements (last visited July 28, 2019). For details on the fee schedule, see *USPTO Fee Schedule*, USPTO, https://www.uspto.gov/learning-and-resources/fees-and-payment/uspto-fee-schedule#TM%20Process%20Fee (last visited April 13, 2020).

192. *Trademark Process*, USPTO, https://www.uspto.gov/trademarks-getting-started/trademark-process#step3 (last visited July 28, 2019).

193. *Search Trademark Database*, USPTO, https://www.uspto.gov/trademarks-application-process/search-trademark-database (last visited July 16, 2018).

194. *Id*. Specifically, the website advises:

> [D]eciding what to search for and interpreting your results can be complicated. There are many factors to consider in determining **likelihood of confusion**. We can't advise you on how to do a clearance search for your mark, do one for you, or interpret your search results. Therefore, we strongly encourage you to **hire a U.S.-licensed attorney** who specializes in trademark law to guide you throughout the application process.

*Id.* (emphasis added).

and, therefore, trademark search engines are likely optimized in a way to ensure that attorneys can trust their results as being definitive. Routinely not returning an accurate result for a potential conflict could cause attorneys to shift business away from one search engine toward another.

Here, as discussed above, private vendors have emerged to assist attorneys and applicants with their own search processes. These trademark search engines all use some form of search algorithm to power their results, although each of them utilizes different methods of integrating data and machine learning into their analytical performance. Again, these can vary in complexity and may be geared toward different audiences. The broad takeaway is that they each represent a means of helping registrants navigate a complicated search problem by reducing search costs for marks through recent advancements in technology. By giving applicants the ability to go beyond what TESS or a library search can provide, they potentially reduce search costs considerably.

The core type of search that each search engine provides is the "knockout search." A knockout search is essentially a trademark search that intends to return marks that are likely to be cited in a 2(d) "likelihood of confusion" rejection for a new trademark application.[195] This category is what we focus most of our empirical analysis on because it is the one point of common ground between all of the search engines in our study. Within the knockout search, there are still some ways that different search engines distinguish themselves. Some may simply reference the USPTO's own TESS search engine,[196] while others combine that data with their own methods.[197] Still others will attach likelihoods for risk scores, which requires a more algorithmic approach than simply checking against TESS.[198]

---

195. *What is a Trademark Knockout Search?*, PAT. TRADEMARK BLOG, http://www.patenttrademarkblog.com/trademark-knockout-search/ (last visited July 28, 2019). According to the Trademark Manual of Examining Procedure, likelihood of confusion refers to a mark that, "as used on or in connection with the specified goods or services, so resembles a *registered* mark as to be likely to cause confusion." TMEP § 1207.01, *available at* https://tmep.uspto.gov/RDMS/TMEP/current#/current/TMEP-1200d1e5044.html.

196. Trademarkia's free service does this, for example. *See* TRADEMARKIA, https://www.trademarkia.com/ (last visited July 28, 2019).

197. *See, e.g.*, TRADEMARKNOW, https://www.trademarknow.com/ (last visited July 28, 2019) (optimizing for speed by prioritizing returning "exact matches"). In its ExaMatch (https://www.trademarknow.com/products/examatch) page, it includes a search engine to search the USPTO and E.U. databases.

198. Both Markify and CSC provide likelihood measures with their results. *See* MARKIFY, https://www.markify.com (last visited July 28, 2019); *see also* CORPORATION SERVICE COMPANY (CSC), https://www.cscglobal.com/global/web/csc//trademark-searching.html (last visited July 28, 2019); *infra* Figure 5.

Different search engines differentiate their core products, so making comparisons between them necessarily simplifies the typical use case for each one. Different search engines will provide different metrics, and there are a few other considerations as well.[199] Many of the search engines in our study distinguish themselves by offering a "comprehensive search" of some sort.[200] These services can vary considerably between different search engines. One major consideration is whether a comprehensive search involves automation or human review. Some comprehensive search tools will automatically generate detailed reports, whereas others have human beings thoroughly investigate a potential mark.

In sum, because of the diversity in trademark search products, evaluating their performance can be tricky. Trademarks have several different elements, and there are multiple ways that a trademark application can be "confusingly similar." Moreover, identifying a "confusingly similar" registration involves some judgment as well because different search engines could return noisier results than others, even if the "correct" answer is present in all of them. Namely, a trademark application can be similar to an existing one in its visuals, phonetics, concept, or spelling.[201] To address this issue, Idan Mosseri and colleagues created "TradeMarker" software, which conducts a variety of independent searches, developing metrics of automated content similarity, image/pixel text similarity, and manual content similarity.[202] They construct individual similarity measures for each of these categories, and then combine each of these measures for an "overall similarity" score.[203] This mixture is useful because it avoids situations where two marks are unlikely to be considered "confusingly similar," even if they share some aspect of their marks. For example, Target and Vodafone have very similar logos, but do not share text, conceptual, or spelling similarities and therefore would not have a high combined similarity score.

---

    199. For instance, whether a conflicting mark is "live" or "dead" is relevant as a dead mark cannot be cited as a reason to reject a proposed mark. *Searching Marks in USPTO Database*, USPTO, https://www.uspto.gov/trademarks-getting-started/trademark-basics/searching-marks-uspto-database (last visited July 28, 2019).
    200. Trademarkia explicitly talks about a comprehensive search, while others like Corsearch offer a "trademark screening platform." *See* TRADEMARKIA, *supra* note 196; *Trademark Screening*, CORSEARCH, INC., https://www.corsearch.com/our-products/trademark-screening/ (last visited July 28, 2019).
    201. *See Possible Grounds for Refusal of a Mark*, USPTO, https://www.uspto.gov/trademark/additional-guidance-and-resources/possible-grounds-refusal-mark (last visited July 28, 2019).
    202. Idan Mosseri, Matan Rusanovsky &Gal Oren, *TradeMarker – Artificial Intelligence Based Trademarks Similarity Search Engine*, SPRINGER NATURE SWITZ. 97 (2019), *available at* https://www.researchgate.net/publication/334352698_TradeMarker_-_Artificial_Intelligence_Based_Trademarks_Similarity_Search_Engine/link/5d2865cd458515c11c27b220/download; *see also* Tim Lince, *How AI will revolutionize trademark searches*, WORLD TRADEMARK REV. (July 2, 2019), https://www.worldtrademarkreview.com/ip-offices/how-ai-will-revolutionise-trademark-searches (highlighting guest analysis provided by TradeMarker that combines visual, semantic/content, and text similarity).

It is also worth noting that each trademark search engine firm offers services beyond just search. Many search engines offer active trademark screening, which takes a client's existing marks and checks to see if potential conflicting marks have been applied for or registered.[204] Again, the USPTO does not take responsibility for enforcing trademarks against potential infringers,[205] and therefore likely created a market for these technologies. This sort of service gives companies the ability to engage in brand management. Brand management is at the core of why trademark law exists and is of high importance to firms, and the searches involved with these activities likely mirror the core technologies powering the core search engine functionality.[206] Below, we outline some of the major characteristics of the trademark search engines we studied.

B.    A COMPARISON OF TRADEMARK SEARCH ENGINES

    1.  *Public Search Engines*

    a)  USPTO

The USPTO offers TESS to search existing trademarks for a potential conflict. TESS allows users to search marks that have been both registered and applied for, but it does not automatically flag conflicts on its own. Instead, the USPTO suggests that users supplement a TESS search by consulting an attorney or using a trademark search firm.[207] TESS further offers a few different options for search inclusiveness, depending on the user's sophistication. Its basic search function does a simple search for word matches, whereas its more advanced engines use design mark codes and other information to construct results.

---

    203.  *See* Mosseri et al., *supra* note 202 ("This separation enables us to benefit from the advantages of each aspect, as opposed to combining them into one similarity aspect and diminishing the significance of each one of them.").
    204.  For example, both Markify and Corsearch offer these services; see descriptions of each search engine below.
    205.  U.S. PAT. & TRADEMARK OFFICE, *supra* note 164.
    206.  *See generally* Landes & Posner, *supra* note 42 (discussing the economics of trademark's signaling quality to consumers).
    207.  *Search Trademark Database, supra* note 193 (see "Trademark Searching" and "Hiring an Attorney").

TESS was launched in 2000, making it one of the oldest systems in our study.[208] At the time it was launched, the USPTO explained that TESS used the same search engine and database that its own examiners use.[209] However, few details are available about the exact search algorithm. One main disadvantage of TESS is that it seems to have relatively few computational resources, as only a fixed number of people may search at once and it requests that users log out to release resources to others in the queue.[210] Previously, the USPTO offered a different free search service since 1998, but TESS ultimately replaced it.

Importantly, TESS also draws from the U.S. government's trademarks dataset.[211] The trademark case files dataset[212] contains information about over eight million trademarks and is the authoritative source for existing and previous trademarks in the United States. The advantage of TESS is that it draws directly upon this dataset, and consequently uses it to generate its own search results.

Although its underlying search algorithm and use of AI is unclear, TESS does have a number of useful features for potential registrants. It provides serial numbers, registration numbers, and whether a conflicting mark is live or dead, like shown in Figure 2. Some ordering occurs as exact matches tend to appear near the top of the search results, but this exact mechanism has not been verified.

---

208. Press Release, *USPTO Introduces New Trademark Electronic Search System*, USPTO (Feb. 29, 2000), https://www.uspto.gov/about-us/news-updates/uspto-introduces-new-trademark-electronic-search-system.
209. *Id.*
210. *Trademark Search: Beginners Guide to Everything to Know*, UPCOUNSEL, INC., https://www.upcounsel.com/trademark-search (last visited July 28, 2019).
211. Stuart J.H. Graham, Galen Hancokc, Alan C. Marco & Amanda Myers, *The USPTO Trademark Case Files Dataset: Descriptions, Lessons, and Insights*, 22 J. ECON. & MGMT. STRATEGY 669 (2013); *see also* Trademark Electronic Search System (TESS), USPTO, http://tmsearch.uspto.gov/bin/gate.exe?f=tess&state=4806:pvkuk8.1.1 (last visited Jan. 23, 2021) ("This search engine allows you to search the USPTO's database of registered trademarks and prior pending applications to find marks that may prevent registration due to a likelihood of confusion refusal.").
212. *See Trademark Case Files Dataset*, USPTO, https://www.uspto.gov/learning-and-resources/electronic-data-products/trademark-case-files-dataset-0 (last visited July 28, 2019).

Figure 2: TESS Search Results

### 2. Private Search Engines

#### a) Corsearch

Corsearch is a relative newcomer to the trademark AI space, having become its own independent company in 2017.[213] It is headquartered in New York City and mainly serves corporate customers, according to Crunchbase.[214] Corsearch is a "brand management" service that offers a range of tools to serve this end. These tools are largely powered by AI and search optimization techniques, and much of Corsearch's value-add seems to be in speed, ease-of-use, and comprehensiveness.[215]

The company offers an array of IP services. Namely, it offers trademark screening, trademark searching, trademark watching, online brand protection, and domain name services.[216] Basically, Corsearch provides a suite of tools for brand management, broadly construed. Its tools allow a user to screen for potential conflicts before filing, search globally once they are ready to do an exhaustive search, watch for potential new conflicts after the mark has been registered, and take legal action against potential infringers.[217] Thus, it creates a complete trademark workflow for potential registrants.

In our study, we focus on Corsearch's "trademark screening" product. The trademark screening engine is a dashboard that provides search results for queries, along with some additional services like visualization, document creation, etc. We focus on trademark screening because it is the closest equivalent to the main trademark search product offered by all of the engines in our study.

That being said, Corsearch claims to distinguish itself from its closest competitors in a number of ways. According to its webpage, its main value lies in its "phonetic search engine." The phonetic search engine allows a user to see results that include phonetic, spelling, and plural variations. Theoretically, this should allow the engine to cover idiosyncratic spellings, and therefore help a client do an exhaustive search for any potential conflicts. Later, we explore phonetic matches as this is one of the common ways a mark application can be rejected,[218] and Corsearch plausibly has a comparative advantage in these sorts of searches.

---

213. *See Corsearch Inc.*, BLOOMBERG L.P., https://www.bloomberg.com/profile/company/1632077D:US (last visited July 28, 2019).
214. *Corsearch*, CRUNCHBASE, INC., https://www.crunchbase.com/organization/corsearch#section-lists-featuring-this-company (last visited July 28, 2019).
215. *Id.*
216. *Our Solutions*, CORSEARCH, INC., https://corsearch.com/our-products/products-overview/ (last visited July 28, 2019).
217. *Id.*

Recently, Corsearch acquired Principium to bolster its own trademark watching services.[219] In addition to its other recent acquisitions, Corsearch is building its portfolio to ensure that it can compete on every aspect of brand management.[220]

**Figure 3: Corsearch's Search Terminal[221]**

**Figure 4: Corsearch Example Results**

---

218. *See Possible Grounds for Refusal of a Mark*, USPTO, https://www.uspto.gov/trademark/additional-guidance-and-resources/possible-grounds-refusal-mark (last visited Jan. 23, 2021); *see also* Beebe & Fromer, *supra* note 56, at 1039 (discussing how the FDA also uses phonetic similarity to determine whether drug names are confusingly similar to one another).

219. *See Corsearch Acquires Principium Trademark Watch and Domain Services Businesses*, BUSINESSWIRE (May 17, 2019, 4:00 AM), https://www.businesswire.com/news/home/20190517005089/en/Corsearch-Acquires-Principium-Trademark-Watch-Domain-Services.

220. *See id.*

221. Note that it includes several language-based search parameters including phonetic search.

b) Markify

Markify was founded in 2009,[222] and is exclusively specialized in trademark searches and brand management. Markify is headquartered in Sweden and provides global services that allow clients to search and manage trademarks across numerous jurisdictions.[223] In 2017, LegalZoom, an American legal technology company, partnered with Markify to power its own trademark and monitoring services.[224] LegalZoom specializes in providing legal help to small businesses and other entities.[225] One of these services is trademark registration, and LegalZoom provides a trademark search as part of its process.[226] Because of LegalZoom's dominance in the U.S. market, its partnership is a key part of Markify's portfolio.[227]

Markify provides several services as part of its general brand management offerings.[228] These include its Comprehensive Search, ProSearch, trademark watch, domain name watch, and an API.[229] The ProSearch search feature is the closest equivalent to other search engines in our study, and thus we focus on this product.[230] The trademark watching service actively checks international trademark databases and provides weekly reports about potential conflicts.[231] Similarly, the domain name watch looks for confusingly similar domain names.[232]

Markify's services are powered by its own trademark similarity search algorithm. The company argues that it distinguishes itself by developing its algorithm from a statistical perspective, so that users can prioritize search results more easily.[233] This approach quite explicitly leverages artificial

---

222. *See Markify*, CRUNCHBASE, https://www.crunchbase.com/organization/markify (last visited Jan. 23, 2021). Please see our first footnote, noting that Markify provided funding for this study.

223. *Id.*

224. *See LegalZoom Selects Markify as Trademark Search and Monitoring Provider*, BUSINESSWIRE (Nov. 07, 2017), https://www.businesswire.com/news/home/20171107005509/en/LegalZoom-Selects-Markify-Trademark-Search-Monitoring-Provider.

225. *Id.*

226. Joe Runge, *Why Do I Need to Conduct a Trademark Search?*, LEGALZOOM.COM, INC., https://www.legalzoom.com/articles/why-do-i-need-to-conduct-a-trademark-search (last visited July 28, 2019).

227. BUSINESSWIRE, *supra* note 224.

228. *See Products & Pricing*, MARKIFY, https://www.markify.com/ (last visited July 28, 2019).

229. *Id.*

230. *Id.*

231. *Id.*

232. *Id.*

233. It says, "The trademark search algorithm was developed by a team of mathematicians, linguists and computer scientists. It was built on a statistical analysis of more intelligence, statistical analysis, and big data to transform trademark search.[234] Markify's central goal is to return as many potential conflicts as possible, but to also filter out as much of the "noise" as possible.[235] Noise in this case would be search results that do not actually present a conflict or are not plausible 2(d) violations.

**Figure 5: Markify Search Results[236]**

---

than 8[,]000 actual cases where a government official had ruled that two trademarks were confusingly similar. The trademark search technology is constantly upgraded and adapted to new markets." *About Markify*, MARKIFY, https://www.markify.com/about.html (last visited July 28, 2019); *see also Big Promises, Big Data*, WORLD INTELL. PROP. REV. (May 21, 2019), https://www.worldipreview.com/contributed-article/big-promises-big-data (noting Markify's role in harnessing big data).

234. *Id.*

235. *See Get a real comprehensive trademark watch service*, MARKIFY, https://www.markify.com/services/trademark-watch.html (last visited Jan. 23, 2021) (discussing its "signal-to-noise ratio").

236. Note that in addition to the mark name, Markify returns a "risk level" that allows users to order results from most to least serious threats to their proposed mark. This image comes from Markify's "comprehensive reports" while we used its "prosearch" for the comparisons between firms.

c) Trademarkia

Trademarkia is a visual trademark search engine that operates as a subsidiary of LegalForce, an intellectual property law firm.[237] Trademarkia offers a number of trademark services including registration, legal action against infringing marks, trademark renewal, trademark revival, and trademark watch.[238] Like other firms in this study, Trademarkia offers several services that can broadly be considered to be "brand management."

Trademarkia distinguishes between "knockout" and "comprehensive" searches and offers both. Knockout searches comb the USPTO page for any similar marks, but these results do not guarantee that the identified marks are available or meet the standard for registrability.[239] Its comprehensive search, on the other hand, furnishes users with a report that checks the mark against additional sources and contexts to ensure that the mark is available. It is a little unclear, but it seems that this process involves human input.

We focus on Trademarkia's knockout searches, although we note that this service is free, and thus Trademarkia's comprehensive search results may be better. However, the free service most closely resembles the other search engines in our study because it appears to use an algorithmic approach without human input.

---

237. *See* LEGALFORCE RAPC WORLDWIDE, https://www.legalforcelaw.com/ (last visited July 28, 2019); *see also Trademarkia*, CRUNCHBASE, INC., https://www.crunchbase.com/organization/trademarkia#section-overview (last visited July 28, 2019).

238. See the "Trademark" drop down menu on https://www.trademarkia.com/.

239. For example, if one searched "google" as a trademark on Trademarkia's free service, a note appears at the bottom of the search results that says:

> **NOTE:** Trademarkia.com is updated regularly with the latest trademarks from the United States Patent & Trademark Office (USPTO). There may be marks that were removed from Trademarkia at mark owner's request. Trademark search results are not indicative of the availability of the trademark. Applications requested through Trademarkia are evaluated by an attorney for the availability of the trademark. The Google trademark has a greater likelihood of registration if it satisfies the following conditions: (1) it is not confusingly similar to other marks, (2) it does not dilute a famous mark, (3) it is not generic or descriptive, and (4) if there are no unregistered, common law trademark holders that are using this trademark in commerce today."

TRADEMARKIA, https://www.trademarkia.com/trademarks-search.aspx?tn=google.

Figure 6: Trademarkia's Search Results[240]

### d) TrademarkNow

TrademarkNow was founded in 2012[241] and is explicitly premised on using AI to revolutionize trademark search. It explains that its search engine,

> [a]t its core is a unique artificial intelligence model of trademark law based on both explicit and intricate domain models of the law. Created by experts in trademark law and linguistics, our cutting-edge system also utilizes state-of-the-art machine-learning techniques to produce models that seamlessly take real-world complexities into account.[242]

Essentially, it tries to encode law, legal rules, and intuitions about legal interpretation of IP into its models in order to furnish users with the most relevant results.

TrademarkNow offers a few different products.[243] ExaMatch is intended to be a first step for any trademark applicant and promises "instant screening" results for trademark results.[244] The company also offers a "clearance search" algorithm called NameCheck, which we used for our analysis, that improves upon knockout searches, and also a brand protection service called NameWatch that checks to see if anyone tries to register a conflicting mark.

---

240. Note that it provides descriptions and statuses in addition to the mark name.
241. *TrademarkNow*, CRUNCHBASE, INC., https://www.crunchbase.com/organization/trademarknow (last visited July 28, 2019).
242. *About Us*, TRADEMARKNOW, https://www.trademarknow.com/about (last visited July 28, 2019).
243. *See Brand Protection – NameWatchTM*, TRADEMARKNOW, https://www.trademarknow.com/products/namewatch (last visited July 28, 2019).

### 3. Our Methodology

Although there are strong theoretical underpinnings for trademark search, there is little systematic evidence about how searches occur in practice. The fact that multiple products exist to assist potential registrants suggests that there is a real demand for tools that ease the trademark search process. In this section, we present results from a novel exploration of the efficacy of various trademark search engines. By comparing and contrasting particular results, we studied how well these search engines identify potential conflicts under Section 2(d) of the Trademark Act, 15 U.S.C. § 1052(d),[245] which forbids the registration of a trademark that is confusingly similar[246] to an existing registered trademark.

As discussed below, answering this broad research question turns on making choices about particular metrics. Our basic approach involved searching across each trademark search engine to evaluate how well each one picks up on potential conflicts. To address this question, we generated a list of "conflicted marks" that we knew should be flagged as a potential 2(d) violation. Second, using this list, we ran searches across each engine, and then measured the returned results. We then compared the results across several search engines using several relevant metrics.

Moreover, in the interest of reproducible research, we also created an end-to-end code pipeline. Each step of the process is entirely programmatic and can be easily reproduced by re-running the same scripts that we ran.[247] The main advantage of taking this approach is that tasks like choosing conflicting marks involved no subjective judgment. Most importantly, automating searches allowed us to conduct this study at scale and collect data that would otherwise take an enormous amount of time and effort to record.[248]

---

244. *Preliminary Trademark Search – ExaMatch*[TM], TRADEMARKNOW, https://www.trademarknow.com/products/examatch (last visited July 28, 2019). TrademarkNow suggests that users, "[s]pend your time on the names that matter and not the ones that don't." *Id.*
245. 15 U.S.C. § 1052 (2018).
246. UPCOUNSEL, INC., s*upra* note 210 (discussing "Likelihood of Confusion FAQ").
247. Our code was mainly written using Python and R, and we will make it available upon request.
248. In this paper, we use about 100 different search terms. In previous work, Moerland and Freitas used terms related to just one mark, "Apple Inc." *See generally* Moerland & Freitas,

To summarize, the basic methodology took the following steps:

1) We developed a list of conflicting proposed marks that should be flagged by a given search engine as being 'confusingly similar' to a preexisting mark.

2) Searched a term across all of the search engines and returned relevant results.

3) Saved all of the search results.

4) Repeated this procedure for each search term.

5) Analyzed the number of killer marks, precision/recall, and other metrics.

### 4. *Generating Conflicted Trademarks*

Generating a list of trademarks to run through search engines was a conceptually challenging task. We wanted to emulate the typical use case as much as possible when doing searches. Tackling this problem meant that the list of trademarks had to resemble actual searches that registrants would reasonably conduct.

One identified potential approach was to take a set of existing registered trademarks, either randomly chosen across goods or services or optimized for a particular trademark class,[249] and search for them in each search engine. The attractive feature of this approach is that searched trademarks should almost definitely be flagged as problematic because there should be an exact match for them in TESS and other databases. If a search engine did not capture this conflict, it would be a signal of poor quality. Unfortunately, searching currently registered trademarks does not reflect how registrants actually use these trademark search engines prior to registration. Since registrants are looking to see whether their own mark is likely to run into a conflict, it is unlikely that anyone would search preexisting marks with any regularity.

Another identified approach is to create fake trademarks to search that closely resemble existing marks. For instance, one could swap a few letters in an existing mark to create a new mark, and then search the new mark to see if it would be flagged as confusingly similar to the original. Again, this approach seems attractive but does not reflect the true data generating process.

---

*supra* note 6. Our method, we think, provides a way to supplement qualitative studies like that of Moerland and Freitas.

249. *See* Brian Farkas, *Trademark Classes: Which One Fits the Mark You Are Registering For?*, NOLO, https://www.nolo.com/legal-encyclopedia/trademark-classes.html (last visited July 28, 2019).

Registrants likely create potential trademarks through creative processes and in ways that are associated with the brand they hope to protect. Swapping out letters and phrases would return marks that are confusingly similar to the originals, but not reflect how registrants actually create their own marks.

For these reasons, we instead scraped recent 2(d) rejections, generated a list of them, and used this list of marks for our searches. This approach overcame the fundamental flaw inherent in other approaches, namely that they do not reflect the actual creative process that generates confusingly similar marks. Additionally, by searching marks that were already rejected for being confusingly similar, we avoided needing to make personal judgments about what the USPTO might consider to be a 2(d) violation. For the same reason, by relying on a list of prior trademarks, we also did not have to occupy the minds of trademark registrants and try to emulate their thought processes when creating trademark names.

Ideally, we would have been able to observe the actual searches that registrants conduct across all search engines. In practice, however, generating this sort of list would be difficult because it would require each search engine firm to disclose its customers' identities, internal algorithms, and business practices in detail. We also had no information regarding whether the marks we searched—or their rejections—were in the datasets that the search engines had been trained on. In machine learning, separating a training set from a validation or test set is important because an algorithm can overfit to the training set, meaning it learns the patterns in that data but does not generalize well. Metrics like accuracy will seem artificially high if reported on the training set for this reason, and therefore a held-out validation/test is important for simulating how the algorithm performs when given *new* data. It is possible that the search engines in our study have seen the marks we search before, but it would be hard to quantify if this happened and whether the results would change.[250] Using 2(d) rejections at least resembles the sort of marks we should expect a search engine to flag, and it avoids the pitfalls of trying to replicate the data generating process wholesale.

---

250. *See* Gareth James, Daniela Witten, Trevor Hastie & Robert Tibshirani, AN INTRODUCTION TO STATISTICAL LEARNING WITH APPLICATIONS IN R 176 (G. Casella, S. Fienberg & I. Olkin eds., 2017), *available at* https://statlearning.com/ISLR%20Seventh%20Printing.pdf [https://web.archive.org/web/20210114184648/https://statlearning.com/ISLR%20Seventh%20Printing.pdf] (explaining this reasoning).

### 5. *Scraping Websites*

Once we generated the list of conflicted trademarks, we turned to running them through each search engine. We wrote Python[251] scripts to achieve this task. Running searches programmatically has several advantages. The primary benefit is that the searches scale easily; conducting ten, a hundred, or a thousand searches requires no additional effort on the part of the analyst, simply more time to run queries. Future studies can therefore use this code as a template to expand upon, confirm, or adjust our results.

Another advantage is the ability to easily make multiple test runs to understand which configurations will get the best results. Search engines have several different search features such as searching for translations, including dead marks, or looking for different types of matches. Optimizing each search engine for its typical use case is key, and being able to run multiple tests easily helps with calibration.

Finally, creating reproducible scripts ensures transparency, which is critical when we are evaluating various software platforms. By enabling anyone to read the code, understand it, and replicate it to guarantee the accuracy of the results, we obviate concerns about mistakes in the research process. These concerns are further mitigated by drawing on the common tools Selenium and BeautifulSoup to complete our research.[252] Combining these two tools make it possible to create scripts that consistently and reliably scrape data from each firm in our study. With relatively simple code, it is possible to generate a rich dataset that allows us to answer a novel research question. Similar studies that look at other areas of law (whether in IP or otherwise) could easily replicate our general approach.

---

251. Python is a popular programming language in software engineering, data science, and other computer programming tasks. It is free to download and use. *See* PYTHON, https://www.python.org/ (last visited July 28, 2019).

252. In terms of technical details, the primary tools we used were the Selenium and BeautifulSoup packages in Python. Selenium is a package that enables automated web browsing through a variety of common browsers. *See* SELENIUM, https://selenium.dev/ (last visited July 28, 2019). Using Selenium, it is possible to automatically navigate to a trademark search website, login, and run search terms. The basic principle underlying Selenium is that if it is possible for a human to click or enter text in any part of a website, it is possible to automate this process with Selenium. The major drawback of Selenium is that if a website's underlying source code changes, then it could potentially break a webcrawler. BeautifulSoup is another common package that can take a webpage and break down its HTML in a convenient format for humans to read. *See Beautiful Soup*, CRUMMY, https://www.crummy.com/software/BeautifulSoup/ (last visited July 28, 2019). The main feature here is that it provides HTML tags for every element on a webpage. In our case, this feature makes it easy to scrape tabular or list results for each of our search terms in an automated fashion.

The pseudocode for accomplishing these steps is as shown in Figure 7.

**Figure 7: Pseudocode**

This program navigates to a trademark search engine, loops through a list of trademarks, searches each mark, and returns a table in a dataframe.

1. Load the list of "conflicted trademarks" to be searched
2. Initialize an empty dataframe object
3. For each trademark in the conflicted trademarks list:
   a. Try:
      i. Initialize a Selenium webdriver
      ii. Navigate to search engine website
      iii. Find the "search bar"
      iv. Enter trademark
      v. Extract XML page source
      vi. Extract table
      vii. Save table to a dataframe
      viii. Click to next page and repeat steps i–vii if necessary to build table sequentially
   b. Except:
      i. Pass

### 6. *Exploratory Data Analysis*

At a basic level, we are interested in whether a search engine provides the user with adequate information to dissuade them from attempting to register a mark that is likely to get 2(d) trademark rejection. However, it is not straightforward to pick a single metric that satisfies this proposition. For this reason, defining metrics is a key task because the core research question could be interpreted in many different ways.

Before turning to an evaluation of the engines, we first provide some exploratory data analysis to build intuition around trademarks, search engines, and the notion of "similarity." For purposes of our study, we identify two major categories of conflicts: either confusingly similar spelling or sound. Since it is easier for a search engine or other algorithm to detect spelling similarity, rather than phonetic similarity, we expect that all search engines will have lower scores on the latter. Still, potential registrants are likely to be concerned with both types of conflicts, thus making the breakdown useful for comparisons between the search engines.

In our exploratory data analysis, we recorded the number of exact matches, as well as the number of phonetic matches. For exact matches, we checked whether the result is exactly the same as the searched mark. For phonetic matches, we employed the Soundex algorithm. The Soundex algorithm matches sounds by taking the first letter of a word, and then encoding the remaining consonants according to a predefined schema that assigns particular letters to particular number values.[253] Note that Soundex is a fairly simple algorithm that is prone to errors, and the search engine's own algorithms are undoubtedly more sophisticated. That being said, we include it mainly to illustrate the concept of phonetic similarity.

We also looked at the total number of results returned by a search engine. Typically, a lawyer is obliged to look through each search result before making a recommendation to a client. Too many results can create unnecessary noise and add to the search costs, but too few results can put the user in the position of mistakenly filing a bad mark. By itself, the number of results is not too informative, but may be useful when put into context with other firms' results. We also provide breakdowns for results tagged as "high" or "low" risk (or the equivalent, whenever available).

a) Baseline

For our baseline, we use the USPTO's TESS results. This choice is a natural baseline because searching TESS is a typical first step for most applicants; it is freely available, and the database is directly connected to the USPTO's own information that it uses in decisions to approve or deny a trademark. By treating the TESS results as a baseline, we implicitly assume that paid services should do better on at least some measures.

The basic results pulled from TESS are in Table 1.

---

253. For more details, see *Soundex System*, NATIONAL ARCHIVES, https://www.archives.gov/research/census/soundex.html (last updated May 30, 2007).

Table 1: TESS Results Example[254]

| marks | conflicted_mark | exact_match | levenshtein_distance |
| --- | --- | --- | --- |
| EX SERIES ZONE | SERIES 1 | FALSE | 8 |
| TOUGHY U-GATE SERIES | SERIES 1 | FALSE | 15 |
| ECONO GATE SERIES | SERIES 1 | FALSE | 12 |
| ATLANTIS G-GUTTER CANOPY SERIES | SERIES 1 | FALSE | 26 |
| KITCHEN ESSENTIALS CLUB SERIES | SERIES 1 | FALSE | 25 |
| EVERY PRIZE ONLY A. AUCTION L THE UNITED | SERIES 1 | FALSE | 254 |
| METAS -SERIES | SERIES 1 | FALSE | 9 |
| LEGACY SERIES | SERIES 1 | FALSE | 9 |
| CHAMPIONS OF EDUCATION MVP SERIES | SERIES 1 | FALSE | 30 |
| MVP SERIES CHAMPIONS OF EDUCATION | SERIES 1 | FALSE | 27 |
| FEDERAL RESERVE BLUNT JMJ THE UNITED | SERIES 1 | FALSE | 265 |
| SQUATTERS SECRET STASH -OFF SERIES | SERIES 1 | FALSE | 28 |
| SYMETRA LINK FIXED INDEX ANNUITY - SERIES | SERIES 1 | FALSE | 36 |
| DR LIONIS: SO NATURAL BUSINESS FIRST AND | SERIES 1 | FALSE | 2070 |
| SERIES | SERIES 1 | FALSE | 3 |

Other search engines might provide even more information than what we study in our paper. For example, whether a mark is live or dead, high/medium/low risk for a violation, and owner information might be provided as well. While this information could be interesting to explore, it is not necessary to answer the core question of how well trademark search engines flag potential 2(d) violations.

Our basic measure for efficacy is the "exact match." An exact match corresponds to an instance where we search a mark that we know was rejected under 2(d), and then see if that exact mark is already registered. In the above example, we looked to see whether the mark "SERIES 1" already exists. An exact match is the most straightforward and least subjective way that a mark can conflict with a preexisting one. The one caveat to this statement is that two marks may share a name if they belong to entirely separate classes, and therefore are unlikely to degrade the quality of the other. In such instances, an exact match does not necessarily result in a rejection.

In terms of exact matches, we show results in Figure 8. In this sample, it is clear that it is fairly uncommon to recover an exact match. About a quarter of the results are exact matches. What makes this figure interesting is that it provides some evidence that trademark search is a more complicated process than simply looking for whether one's proposed mark already exists. Rather, most potential conflicts will not match exactly and therefore require some judgment about likelihood of confusion.

---

254. Here, we used the term "marks" to correspond to a returned result; "conflicted_mark" to refer to a mark that was previously rejected under 2(d); "exact_match" denotes whether, for a given row, the value in "marks" corresponds to the value in "conflicted_mark." "Levenshtein_distance" refers to the edit distance between "marks" and "conflicted_mark." We calculated exact_match and levensthein_distance columns ourselves. The basic procedure is that we would take each of the "conflicted_mark" values (like SERIES 1), search them through each search engine, then store all of the search results as "marks," and calculate these measures.

**Figure 8: Exact Matches in TESS**

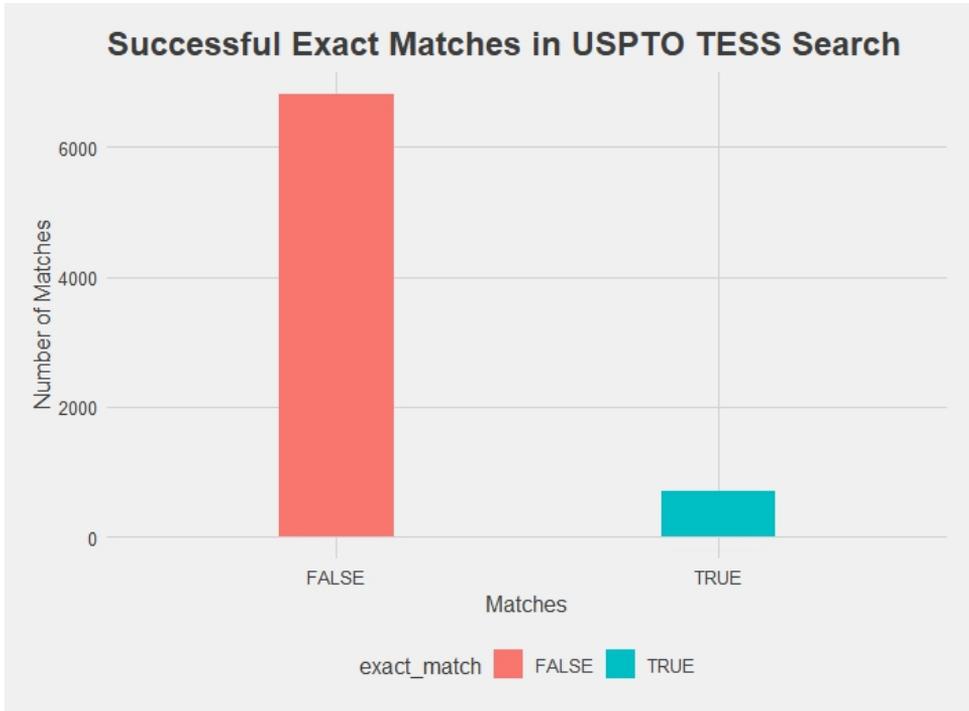

Our next measure was for "phonetic matches," as shown in Figure 9. Using the Soundex algorithm,[255] we looked for whether two strings match based on a phonetic encoding. These results provided more matches, with about half of the results returning a positive phonetic match. Again, the interesting bit here lies in the results that were not matches; discerning whether there is a signal in this group of marks is a key task for any improvements over TESS.

---

255. *Soundex System*, *supra* note 253.

Figure 9: Phonetic Matches in TESS

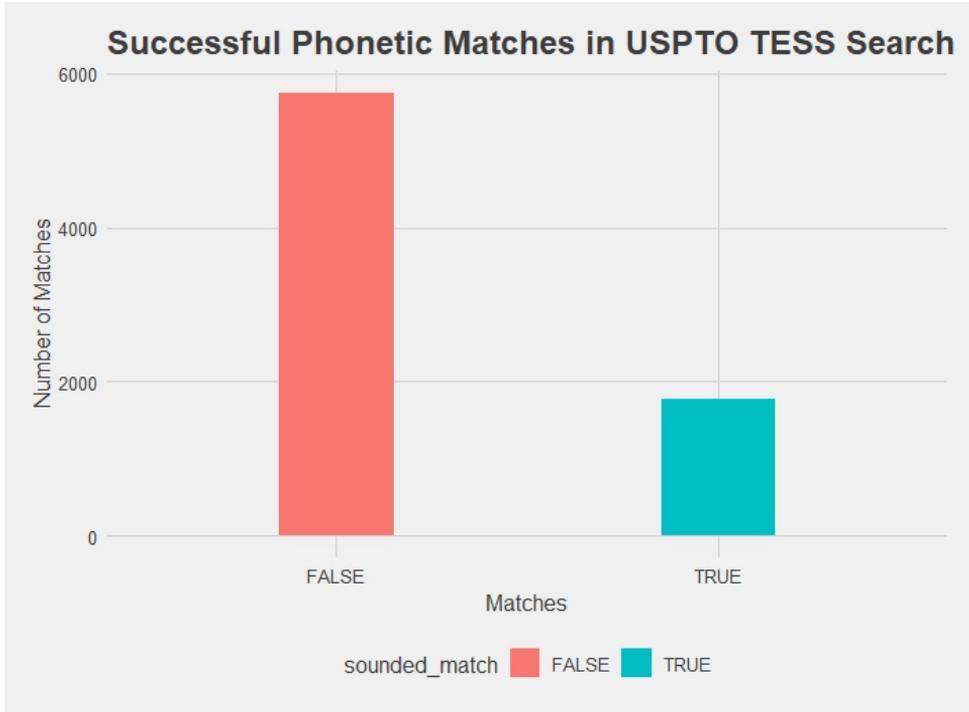

Finally, in Figure 10, we look at the overall number of results returned for each searched mark. This is an important metric because it contains a few key pieces of information. A large number of results could imply that a search engine did a good job exhausting all possible conflicts and returning a lot of relevant information. On the other hand, a large number of results could also imply that a search engine produced a lot of noise, perhaps too much for a human to reasonably sift through. Below, we visualize a random sample of searched marks and the number of results returned for each mark. Again, this is a random sample so one should not draw an inference from the shape of the distribution. However, it does provide a useful baseline for what a registrant can expect to find when they search TESS.

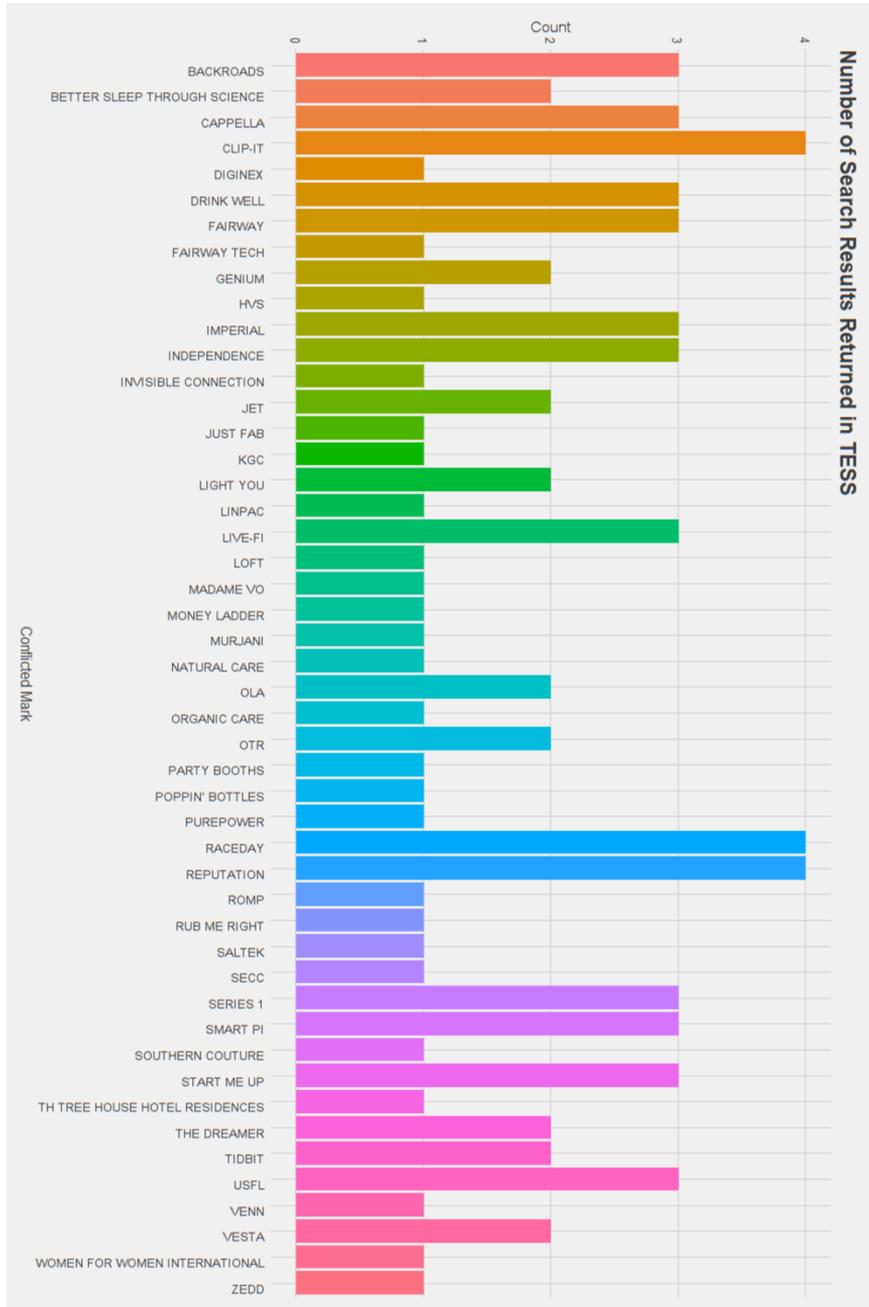

Figure 10: Sample of Number of Returned Search Results Per Mark

b) Exploratory Analysis of Private Search Engines

Below, we present results that compare how AI-powered search engines fare compared to the USPTO's own TESS system. Like above, our basic metrics are number of search results returned, how many exact matches are returned, how many phonetic matches are returned, and how many "close" matches are returned by letter substitution.

The overall takeaway from the results is that AI-powered trademark search engines indeed provide valuable insights for potential applicants. Either by pulling in additional information or packaging it in more manageable ways, they, in general, improve over the baseline results in interesting ways. It is also worth noting that they optimize for different things and thus may be better suited to different use cases.

Consider, for example, the number of matches that each search engine returns. Figure 11 illustrates this point. We searched 115 different marks, and Markify returned around 27,000 potential matches across these, while TESS and Trademarkia returned about 8,000, and TrademarkNow returned approximately 3,000. This comports with expectations, since Trademarkia and TrademarkNow seem to heavily rely on cross-checking against TESS, while Markify pulls in additional sources.[256]

---

256. MARKIFY, *About Markify*, *supra* note 233.

Figure 11: Number of Matches by Search Engine

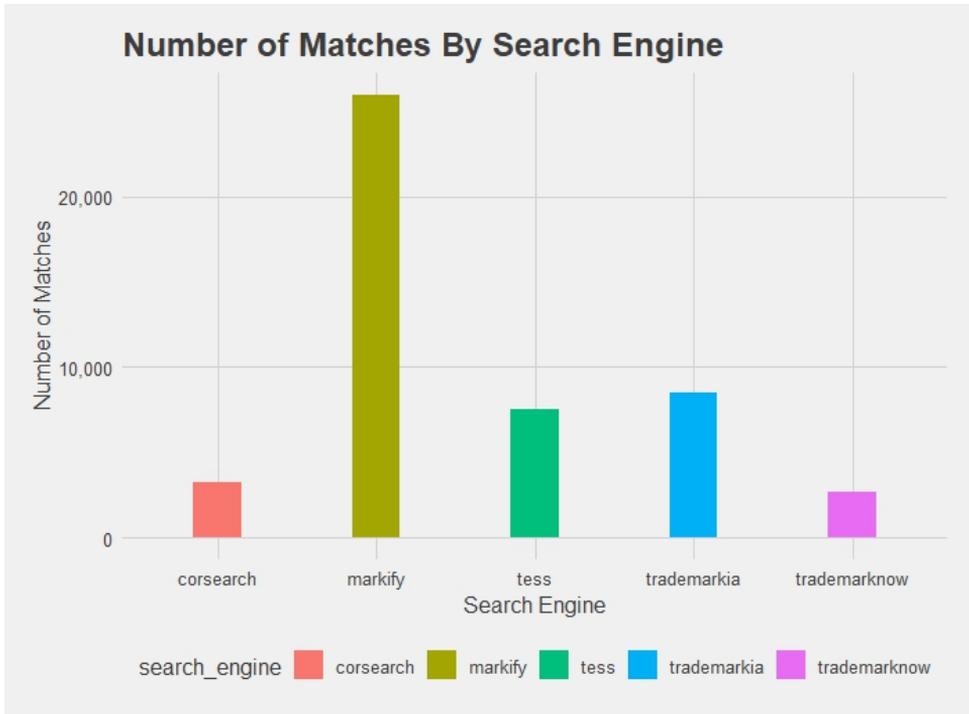

Digging deeper, we can also see this difference visualized across different search terms. Figures 12–16 illustrate the number of search results per mark for each private search engine in our study. Note for these figures, we sampled ten marks to visualize the data. In general, each engine that returned a similar number of relevant results, often between 10 and 20. Certain marks, however, returned a much larger number of results.

Figure 12: Search Results by Search Term in TESS

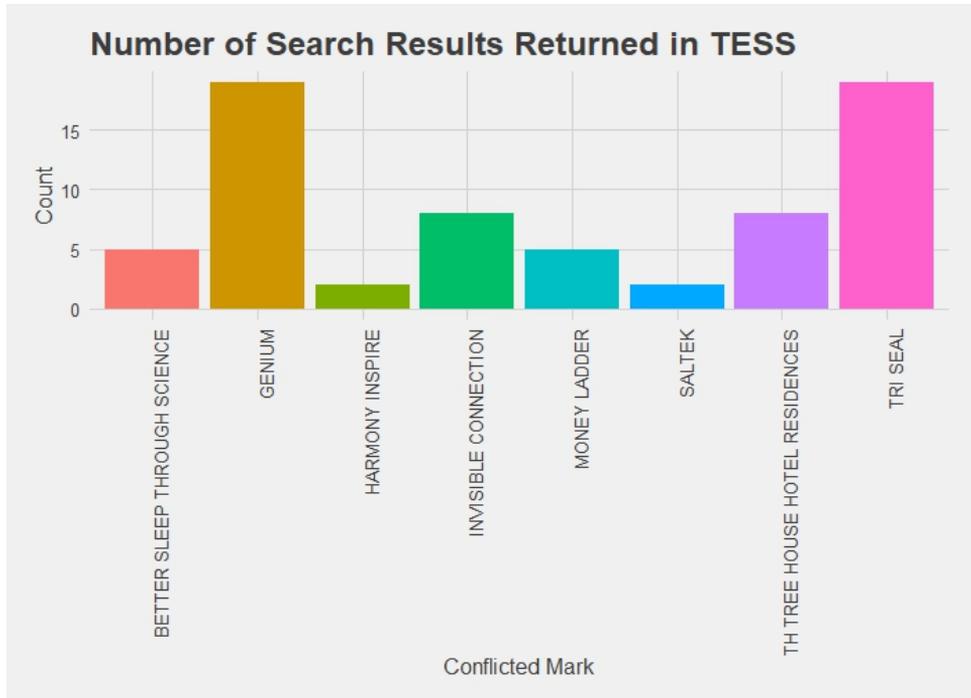

Figure 13: Search Results by Search Term in Markify

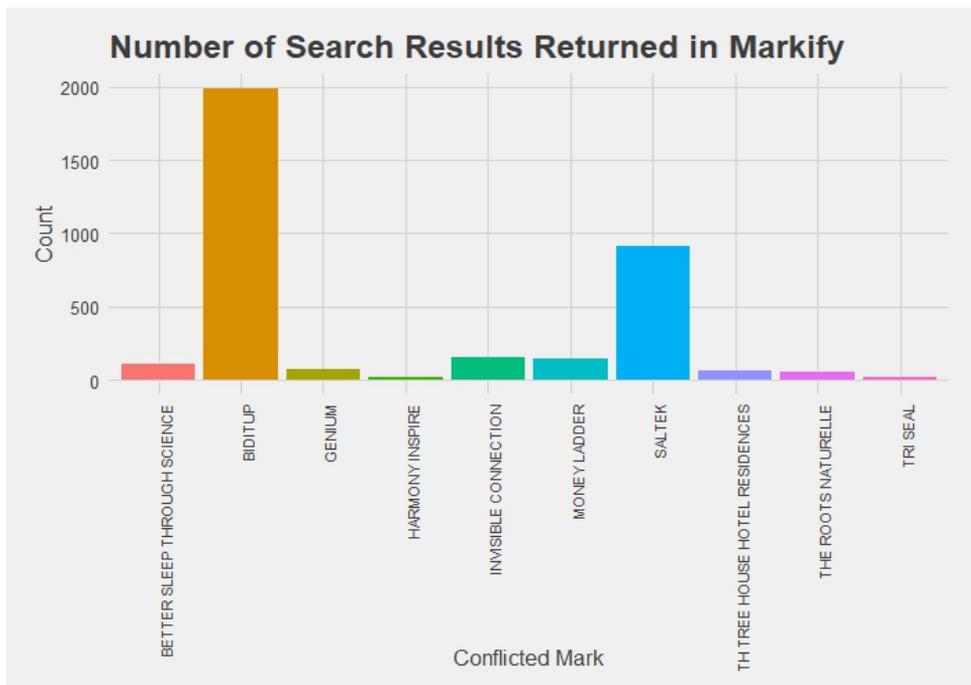

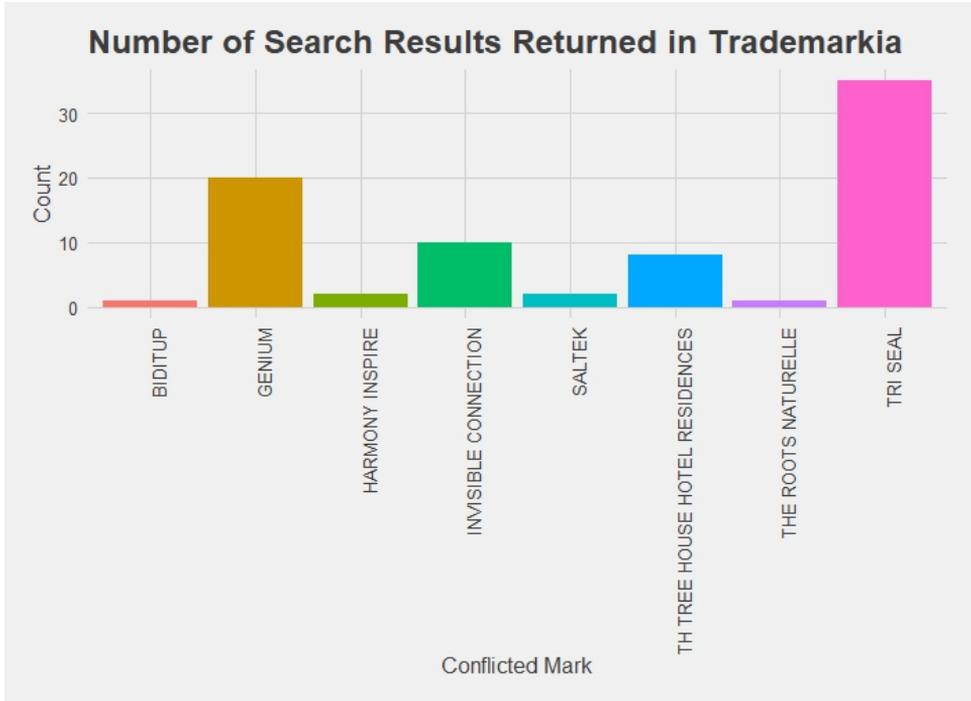

Figure 14: Search Results by Search Term in Trademarkia

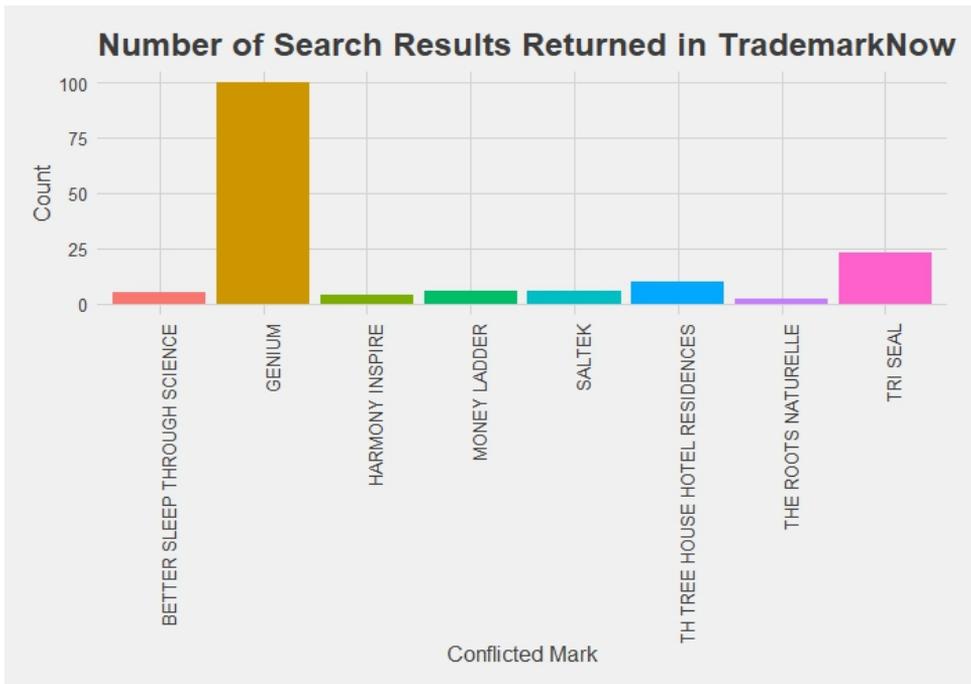

Figure 15: Search Results by Search Term in TrademarkNow

Figure 16: Search Results by Search Term in Corsearch

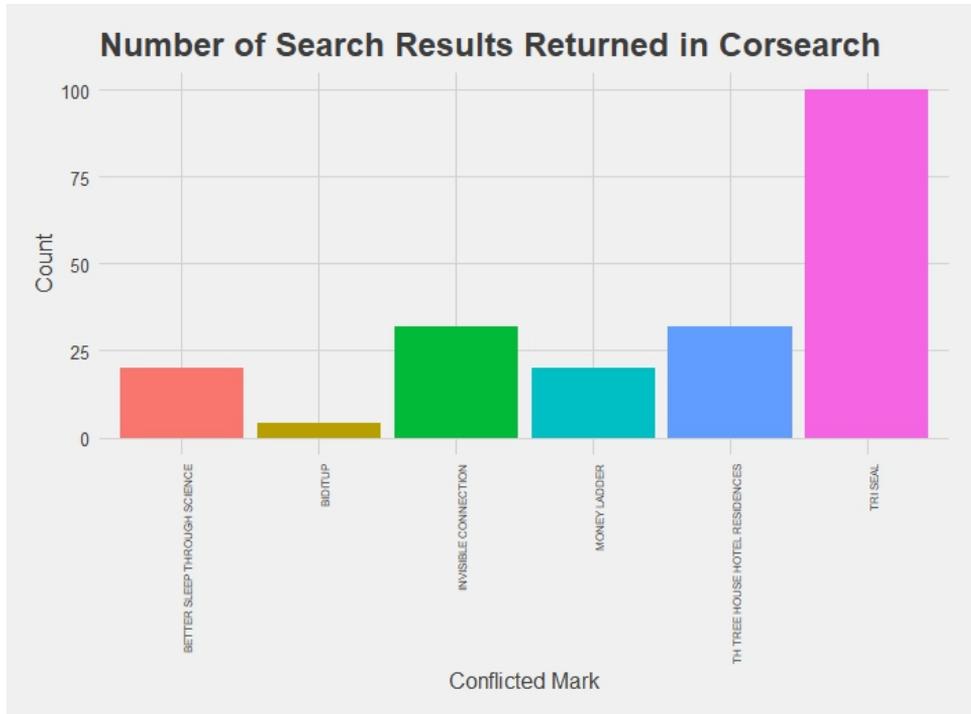

Investigating further, we can see the utility of AI-driven search by looking at our simplest metric, exact matches. TESS, Corsearch, Trademarkia, and TrademarkNow all returned a similar number of exact matches across all searches. However, if a trademark applicant was optimizing solely on finding exact matches, they might prefer a private search engine. Note that in Figure 17, TESS returned many more results that are not exact matches, while Trademarkia and, especially, TrademarkNow filtered out much of this noise. The AI systems underlying both of these private search engines aim to return fewer results overall, and thus better amplify the signal provided by the actual "exact matches" in the data.

Figure 17: Exact Matches in TESS, Trademarkia, and TrademarkNow

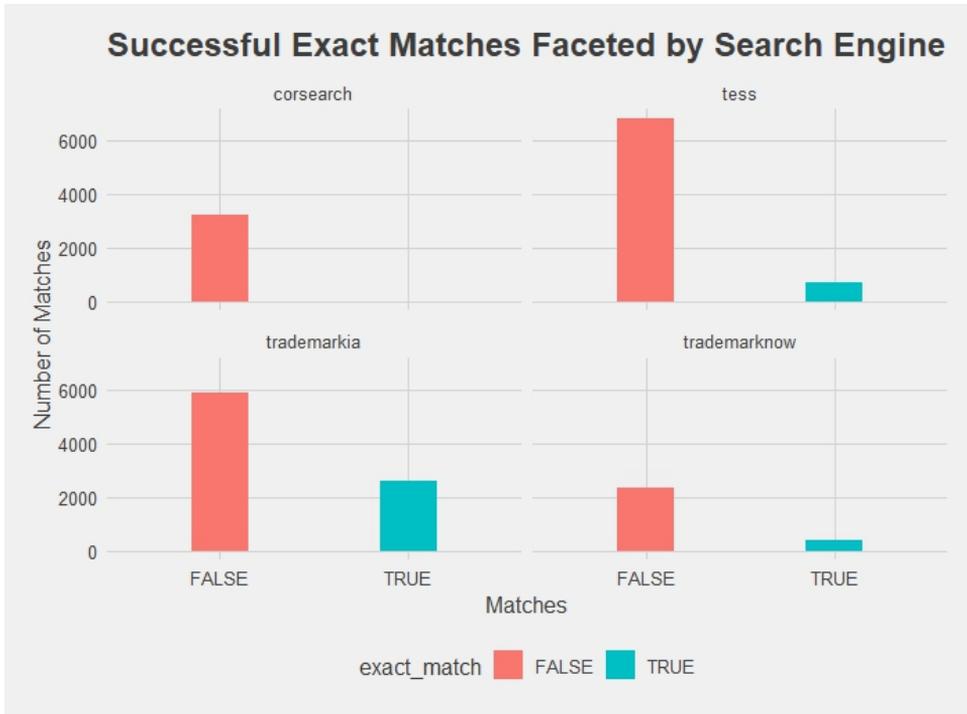

We see a similar phenomenon with phonetic matches as well. Note that in Figure 18, each search engine tends to recover more phonetic matches than exact matches. This result is unsurprising as phonetic matching is less restrictive than exact matching. Also note that Trademarkia does remarkably well as nearly 40% of its results actually match phonetically. Similarly, TrademarkNow achieves nearly 50%. Relative to the TESS baseline, these AI-driven results represent a substantial reduction in noise.

Interestingly, Corsearch, which specializes in phonetic searches, seems to achieve a great deal of noise reduction. It returns fewer results overall than TESS, Trademarkia, and TrademarkNow, but generally returns a higher proportion of phonetic matches. This implies that the algorithm filters out a lot of irrelevant results and does a fairly good job prioritizing actual phonetic matches. Again, Corsearch's own phonetic match algorithm may differ from the Soundex algorithm's rules, so our results may understate the extent to which it successfully finds phonetic matches.

Figure 18: Successful Phonetic Matches Across Search Engines

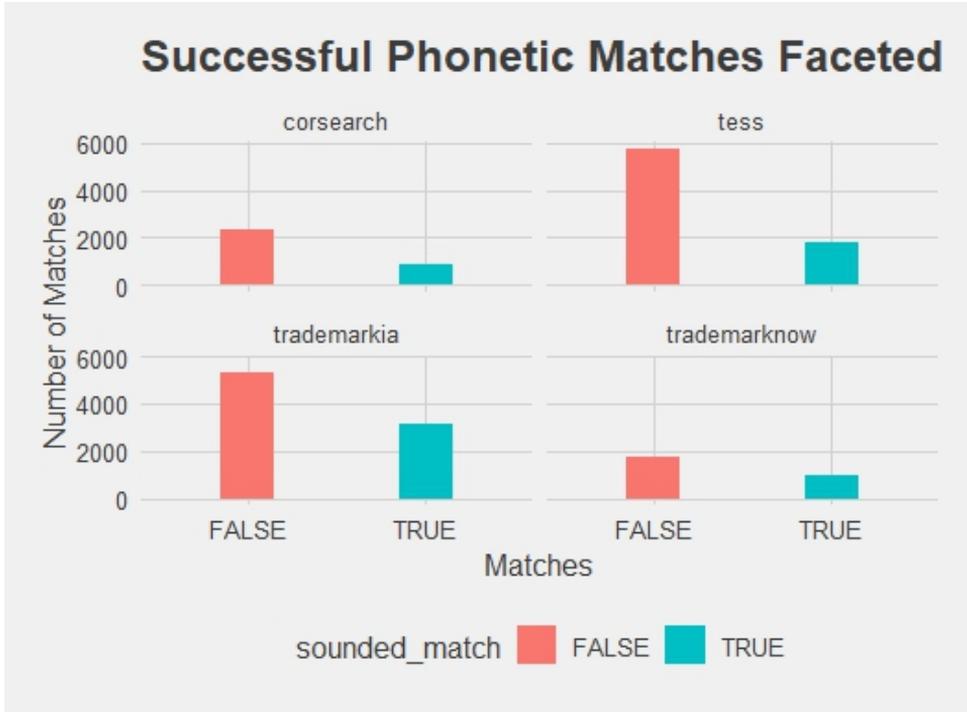

The major takeaway from these results is that AI truly is transforming the trademark search landscape. Even on these basic metrics of exact and phonetic matches, a trademark applicant has little reason to use TESS over a private competitor, particularly when some of these private search engines offer their basic search functions for free (and charge for brand management instead), offering substantial efficiency gains. By returning fewer results in general and successfully filtering out irrelevant results, they make it easier to find knockout conflicts. Basically, this algorithmic approach achieves significant noise reduction at virtually no additional cost to the user.

### 7. Metrics

For our main results, we focused on whether a search engine successfully finds the mark that the USPTO cited in its 2(d) rejection (i.e., the "killer mark"). A killer mark is essentially an existing trademark that justifies rejecting a new application. If a search engine successfully uncovers such a mark, then it succeeded in providing the applicant with information about whether their proposed mark will be accepted. If the search engine fails to find this killer mark, the probability that an applicant goes ahead with a frivolous application rises.

To examine whether the search engines in our study successfully find the killer marks, we used the following metrics. For any given search result, we checked:

**True Positive**: The search result matched a killer mark

**False Positive:** The search result did not correspond to a killer mark

**False Negative:** There was a killer mark that did not have a match in the search results.

Conceptually, these metrics are usually presented alongside "True Negatives." However, we cannot identify true negatives in this context because that would correspond to no search results returned and no killer marks present. That being said, we can still further combine the preceding metrics in useful ways:

**Recall:** Ratio of killer marks found to total killer marks, i.e., True Positive/ (True Positive + False Negative)

**Precision:** Ratio of search results that were actually killer marks, i.e., True Positive/(True Positives + False Positives).

These metrics are frequently used in machine learning for classification problems and work well in this context, too, because they can give us a sense of how each search engine performs, and the tradeoffs among them. For instance, one search engine may prioritize recall (i.e., finding all of the relevant killer marks) over precision (i.e., not flagging false positives), or vice versa.

In the results section, we present these metrics in a few different ways. First, we tweak these definitions slightly to see how well each search engine does at finding any killer mark (instead of all of the killer marks). We then show precision and recall for all search results in our overall dataset. Finally, we show the same metrics for when we limit the number of returned search results per trademark application. Note that in calculating these numbers, we only used results from each search engine's basic search that was the equivalent of a "knockout" search.

### 8. Results

Our results suggest that the landscape of trademark search is rich and interesting, and there is a real potential to further study search costs borne by trademark registrants. The main takeaway is that private trademark search engines provide a genuine value-add to a potential trademark registrant. While not all private search engines provide a meaningful improvement over free, public options, there is evidence of meaningful differentiation between various products.

Our exploratory analysis illustrates some of the basic questions in trademark search. Specifically, we showed that there is some evidence of differentiation between different search engines and the USPTO's own search engine. Differences in number of results returned, the types of matches, and other features may be relevant. In this part, we look at how each search engine performs with respect to our precision and recall metrics to examine these differences in greater depth.

Before delving directly into precision and recall, we first look at whether a search engine finds at least one killer mark associated with a particular trademark search. Figure 19 shows the number of instances in which a search engine finds at least one killer mark. In this case, TESS actually does not perform so poorly relative to private sector search engines. In general, most search engines fail to find a killer mark more often than not.

Figure 19: Killer Marks Found by Search Engine

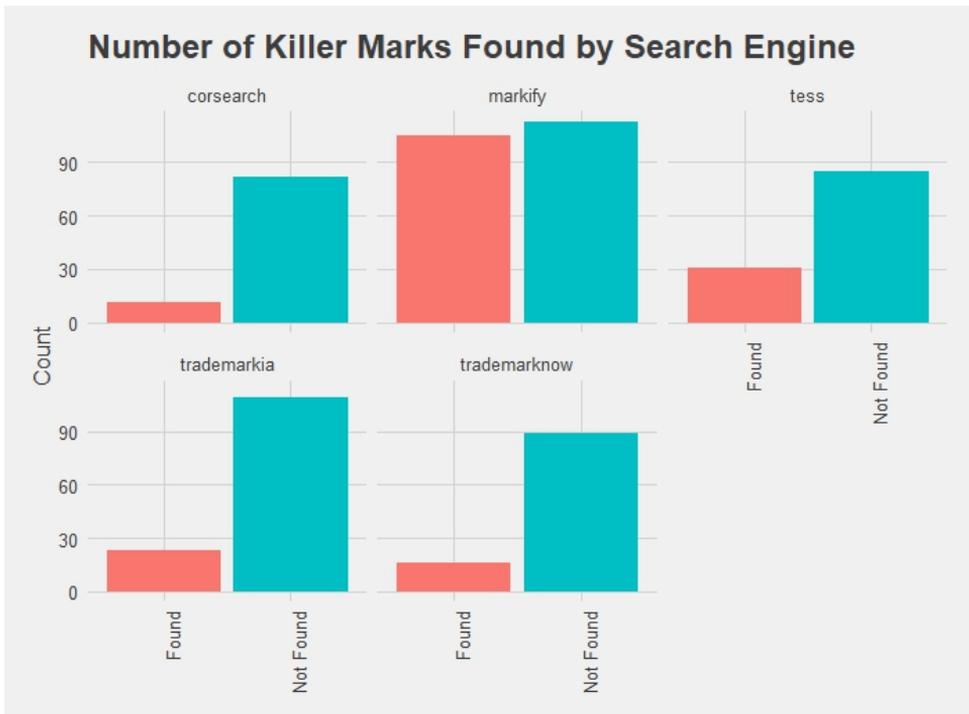

However, we calculate precision and recall somewhat differently. Instead of asking whether a search engine finds at least one killer mark, we instead ask,

"Of all of the killer marks in the dataset, how many was each search engine able to detect?" Some searched trademarks have multiple killer marks associated with them, so precision and recall here will capture whether a search engine uncovered all of the relevant killer marks.

In Table 2, we show results where we derive the precision and recall for each search engine, without limiting the number of results that each search engine returns. Results show that every private search engine achieves higher recall than TESS, and many improve on precision as well.

Table 2: Precision-Recall of Search Engines

| Search Engine | Recall | Precision |
| --- | ---: | ---: |
| Corsearch | 0.369919 | 0.028086 |
| Markify | 0.609756 | 0.006916 |
| Tess | 0.146341 | 0.006524 |
| Trademarkia | 0.105691 | 0.04878 |
| TrademarkNow | 0.691057 | 0.06308 |

We can also explore how a potential applicant could tradeoff between recall and precision. Figure 20 shows the information from Table 2 as a scatter plot with recall on the y-axis, and precision on the x-axis. Figure 21 shows the same information, but this time with varying limits on the number and results per search for each search engine.

Figure 20: Precision and Recall

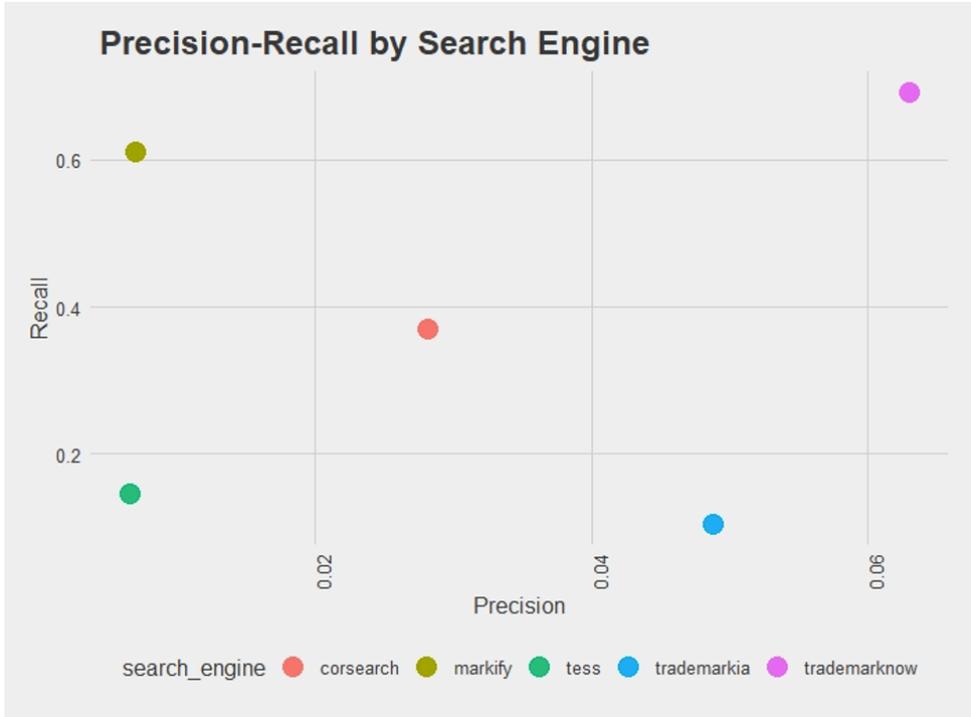

Figure 21: Precision and Recall with Limited Search Results

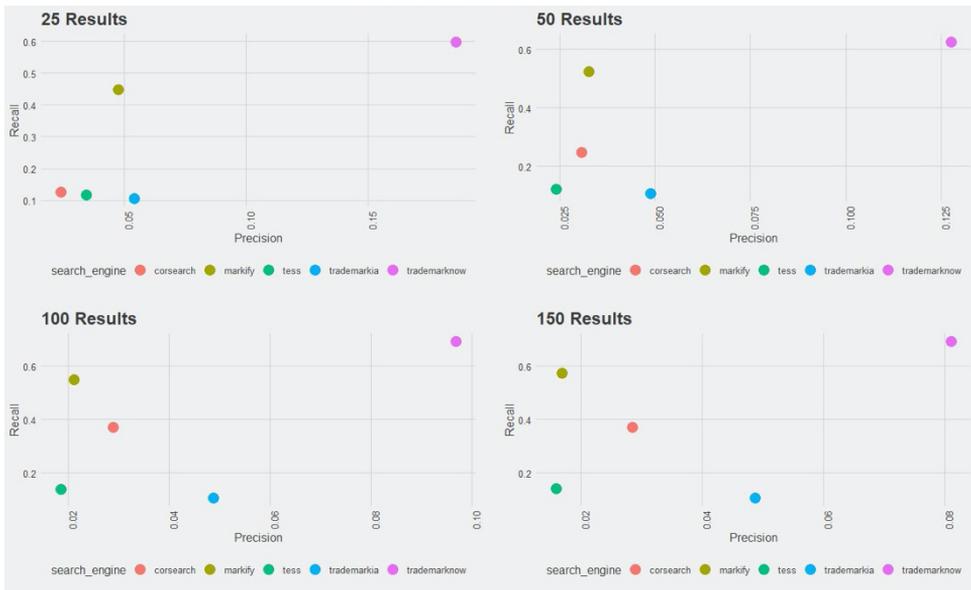

Interestingly, different search engines exhibit different behaviors depending on the limit on the number of search results. Corsearch's precision tends to improve with additional results, while its recall becomes stable around 0.40. Trademarkia also tends to improve on both measures with additional results, and tops out on recall around 0.20. Markify consistently achieves recall in the 0.55–0.60 range, but loses precision with additional results. TrademarkNow similarly achieves a similar recall across the board, but maintains a higher precision than any of the other search engines, even though it decreases with additional searches. More simply, one could stop searching after fifty results and likely already have found the killer mark in Markify and TrademarkNow, while that threshold may be closer to a hundred in Corsearch and Trademarkia. These numbers should also be properly understood as an estimate taken from a sample in a particular time. The marks were scraped in 2019 and searched between 2019 and 2020. Since these are dynamic systems, differences in the sample or time could change these findings considerably.

Taken together, these results indicate that the search engines prioritize certain marks in their search results. Some like Markify and TrademarkNow make this explicit with riskiness indicators.[257] Others seem to do such ordering more implicitly. If precision and recall both increase with additional searches, that indicates that killer marks tend to be identified somewhere other than the beginning of a search result list. On the other hand, if recall remains stable and precision decreases, that indicates that the search engine already found the relevant killer mark.

Both the number of times that a search engine finds a killer mark and the precision and recall scores yield valuable insights. Even one killer mark would be enough to defeat a trademark application, so successfully finding at least one is important for assisting potential trademark registrants. Most trademark search engines do about the same as the USPTO on this measure, and all search engines perform better than the USPTO at finding all possible conflicts.

## IV.  IMPLICATIONS FOR FURTHER STUDY

Today, AI is rapidly reinventing the process of search altogether, particularly in areas of law and government. Court cases, congressional hearings, and government documents are all examples of areas where AI may soon be used in search tools.[258] Already, AI is being deployed to search databases of parking tickets for those who want to contest them.[259] Within the world of IP, we see AI-related techniques throughout the global marketplace, and more and more countries and companies have turned to the tools of machine learning to refine their techniques.

---

   257. Markify uses "high risk" and "low risk" classifiers, while TrademarkNow shows the percentage likelihood of riskiness.
   258. Faraz Dadgostari, Mauricio Guim, Peter A. Beling, Michael A. Livermore & Daniel N. Rockmore, *Modeling Law Search as Prediction*, ARTIFICIAL INTELL. L. (2020), https://

These AI-powered techniques are especially important, not just for the purposes of refining search, but also because of the insights they offer into the economics of IP. On a scholarly level, as our comparison shows, a new area emerges for future research on firm search costs within the trademark registration system through the intersection of AI and trademark search processes. In this Article, we showed how AI is revolutionizing the economics of search in the trademark space, raising new questions about the role of AI in brand management more generally. The main implication of our research is that search costs and AI will continue to be important to legal decisions, both within IP and outside of it. As we showed, firm search costs are a dramatically overlooked area of study and may ultimately hold the key to studying the role of AI in trademark law.

A. OUTCOMES AND IMPLICATIONS

As we have suggested, when scholars and practitioners explore the potential role of AI in transforming patent prosecution and litigation, they may also benefit from looking at trademarks. Trademarks are incredibly valuable assets, and studying their role in the AI-powered marketplace reveals core insights into the economics of IP system at large. As we have shown, AI carries the ability to efficiently compare a proposed trademark against millions of registered trademarks and to assist in determinations about the proposed trademark's worthiness of protection. As with patent and copyright infringement, effective deployment of AI tools prior to the creation of a property right in the trademark context could substantially reduce litigation and other costs when real conflicts arise later on.

At the outset, our legal system places the core responsibility for trademark search and enforcement on the trademark holder.[260] Thus, one interesting question that may be worth exploring is how the use of AI in the USPTO context compares to other government contexts. As we have noted throughout this piece, trademark search engines largely emerged because the USPTO does not enforce existing trademarks against potential conflicts.

---

doi.org/10.1007/s10506-020-09261-5 (suggesting a model of law search based on a notion of search space and search strategies).

259. Shannon Liao, *"World's First Robot Lawyer" Now Available in All 50 States*, THE VERGE, (July 12, 2017), https://www.theverge.com/2017/7/12/15960080/chatbot-ai-legal-donotpay-us-uk.

260. USPTO, *supra* note 188 ("Trademark Basics").

Trademark owners are responsible for discovering and taking legal action against potentially damaging marks.[261]

As such, optimizing the search process benefits both the trademark holder as well as the overall marketplace for trademarks generally, ultimately benefiting consumers. If the USPTO grants too many confusing trademarks, then the market would produce weaker trademarks, harming consumers in the marketplace and leaving more marks vulnerable to enforcement by others. From a registrant's perspective, avoiding a potential rejection saves a lot of time and effort that would otherwise be wasted, conserving the strength of the mark that is ultimately registered. As we have argued, the crucial moment of initial search is a key part of the brand-creation and management process, forming an important threshold of protection.

In this study, we looked at one aspect of the trademarking process—the search for possible conflicts prior to registration—and the significance of search in terms of rethinking our approach to trademark law altogether. The trademark search technologies that we studied here are some examples of how new computational techniques are attempting to solve this puzzle by modeling human decision-making. To summarize, AI lowers search costs by doing a lot of the hard work of making substantive inferences about the relationships between different trademarks, thus empowering applicants to make informed decisions about whether to proceed with their trademark applications. Engstrom et. al. provide an in-depth look at the USPTO's current experiments with AI adjudication, specifically in the realm of patent examination.[262] They note that AI has the potential to reduce search costs for the examiners, but thus far has not been fully implemented as the tools mostly improved the work of examiners with computer science backgrounds.[263] Our study suggests that the development of private sector alternatives in the trademark space might make these tools more broadly accessible. Indeed, the UPSTO is currently exploring implementing deep learning models on image searches.[264]

Ultimately, as we suggest below, our exploration of trademark search engines and the choices we made with regards to methodology and metrics could have interesting lessons for other similar studies in different areas of law. Our study also revealed some important conclusions about the process of trademark registration and the important role that search costs can play in the process.

---

261. USPTO, *supra* note 164.
262. *See generally* Engstrom et al., *supra* note 64.
263. *See generally id.*
264. *See generally id.*

The first main takeaway from this exploration is that AI is already being used in this space, and it is capable of reducing search costs through algorithmically driven information retrieval. Noise reduction and algorithmic prioritization are two major features that these AI search engines achieve. Trademark applicants now have access to tools that can process millions of preexisting trademarks, analyze them, and produce relevant outputs that human beings can understand.

Second, consistent with the literature that finds that consumers give significant weight to non-monetary attributes (like brands, reputation, service quality and pricing quality) in making purchase decisions,[265] we found that trademark registrants, in using AI-powered search, also enlist a variety of non-monetary variables in their own considerations, such as trademark class and lexical similarity to existing marks. This means that search engines can optimize on many more variables than just trademark strength alone. At the same time, it is reasonable to presume that the addition of non-monetary elements, such as the ones that we have seen, can play a determinative role in the trademark registrant's selection of a search engine. Some of these non-monetary attributes may turn on the risk of litigation, the magnetism of the mark, or the mark's relationship to other identities and marks, among others.

A third takeaway involves optimizing the prediction of the outcomes of both registration and potentially litigation. Our results provide some basic validation of the central premise that these types of legal outcomes can be mathematically modeled. These models can detect lexically and phonetically similar marks and, importantly, can sift out results that do not meet certain similarity thresholds. Some attach explicit risk scores, while others implicitly calculate them and then order results. As expected, these risk determinations may follow expected patterns both in distributional shapes and over time, but the patterns may change in the future as adversarial models develop.

None of the trademark search engines we studied model whether a mark objectively meets or fails to meet the 2(d) standard. Such an objective truth plainly does not exist. Rather, these search engines attempt to model the ways that the trademark office, or rather the people in the trademark office, reach their determinations. Implicitly, by making choices about which marks to return and ordering them in a specific way, these search engines make the claim that they can approximate trademark examiners' decision-making well enough to guide trademark applicants' and registrants' business decisions.

Finally, it bears mentioning that including a selection of a larger number of AI-driven variables in a trademark selection decision also introduces the potential for an extremely complex decision-making process. Conceptualizing the potentially unlimited set of variables is practically impossible. The scale of the trademark search space is also massive with millions of registered trademarks in existence. Traversing this massive set of trademarks and retrieving the ones that could present potential conflicts implies enormous search costs for registrants and trademark examiners alike. Search engines try to ameliorate these costs by reducing noise. But, as we have shown, some search engines are better than others at reducing the level of noise encountered by applicants and correcting for the information asymmetries that arise. Specifically, these search engines optimize on certain similarity metrics and drive their results with them. Today, it is difficult to surmise how these AI-driven effects might play out in trademark litigation, i.e., whether they would increase or decrease the occurrence of litigation or its costs. On this point, more research will be needed in the future.

---

265. *See* Zhang et al., *supra* note 23, at 91.

B.   FRAMING TRADEMARK REGISTRATION AS AN ADVERSARIAL MACHINE LEARNING PROBLEM

As these AI search tools mature, we expect that trademark registration will start to resemble an "adversarial machine learning" problem.[266] Previous literature in IP and administrative law identified the back-and-forth between the USPTO and patent applicants.[267] These pieces discussed the problem of the PTO adapting to increasing sophistication in patent applications.[268] This sophistication is in part driven by the use of AI tools, and, in turn, the USPTO might consider using machine learning to improve its own capacity to conduct meaningful examinations.[269] Because applicants have strong incentives to maximize the scope of their claims and the USPTO has an incentive to minimize this scope, the two sides will each construct their decisions in anticipation of the other's incentives.[270]

To build on this literature, we suggest also reframing trademark search as an adversarial machine learning problem. Adversarial machine learning refers to machine learning applications where underlying data distributions change in response to external stimuli. For instance, one problem in training AI for self-driving vehicles is that these AI systems can be easily tricked with just a little additional noise.[271] A self-driving vehicle may be trained to recognize a stop sign with high accuracy, but may suddenly fail if a stop sign has a sticker on it. Although a human being would still recognize the stop sign as such, the AI can be easily fooled because it has never seen this sort of example before.

---

266. For background on adversarial machine learning, see generally Ling Huang, Anthony D. Joseph, Blaine Nelson, Benjamin I.P. Rubinstein & J. D. Tygar, *Adversarial Machine Learning*, AISEC '11 (Oct. 2011), https://dl.acm.org/doi/pdf/10.1145/2046684.2046692.
267. *See generally* Rai, *supra* note 3; Ebrahim et al., *supra* note 3, at 1193–95 (describing the inventor-examiner interaction).
268. *See generally* Rai, *supra* note 3; Ebrahim, *supra* note 3, at 1195–1211 (discussing the automation applications in patent prosecution).
269. *See generally* Huang et al., *supra* note 264.
270. *See generally* Ebrahim, *supra* note 3.

To address this problem, an analyst may try to present the AI with "adversarial" examples in the training phase so that it can learn from these examples. In the self-driving vehicle example, this process could involve perturbing pixels in an image or providing examples of stop signs with stickers and other idiosyncratic markings. Thus, the AI can learn to improve its predictions, even when there is noise present.

Extending this concept into the trademark space, we can conceptualize the general problem articulated by authors such as Rai and Ebrahim in these terms. Consider the following theoretical model: Assume there was a universe of trademark applications prior to the advent of private trademark search engines. Once AI trademark search tools were built based on historical PTO decision data, the recommendations produced by these tools likely influence the names and types of marks in applications to the PTO, thus changing the underlying distribution of trademark applications.[272] The PTO, in response to this change, adjusts its own algorithms and procedures. The search engines retrain their models based on new PTO decisions, and, once again, influence the sorts of trademark applications that are eventually filed. And the PTO again must update its decision-making. This interplay between the PTO and trademark search engines (and trademark applicants) thus evolves dynamically over time.

By framing trademark registration as an adversarial machine learning problem, it becomes clear that the introduction of AI into the process of trademark registration also changes the substance of trademarks. When the PTO makes a series of decisions that search engines must retrain their models on, this represents the PTO adding new noise into their systems. Similarly, when trademark applicants file new applications that are optimized by advice provided by search engines, they add noise to the PTO's decision-making. This dynamic game implies that, over time, the substance of applied for and registered trademarks may keep changing.

---

271. Solving the problem of malicious signage in particular is an active area of research in computer science. *See generally* Chawin Sitawarin, Arjun Nitin Bhagoji, Arsalan Mosenia, Mung Chiang & Prateek Mittal, *DARTS: Deceiving Autonomous Cars with Toxic Signs*, ARXIV.ORG (May 31, 2018), https://arxiv.org/pdf/1802.06430.pdf.

272. Indeed, we mention these selection effects as a hurdle for studying the causal effect of trademark searches in our methodology section.

Previous scholars have advocated for the use of machine learning in patent examinations as a response to increasing sophistication in the private sector. Adversarial machine learning makes clear why this call is important. The term "adversarial" may imply that the contest between the PTO and private sector search firms is damaging, but it should instead be thought of as a framework that improves the quality of trademarks and administrative decision-making. Here, the outputs of the PTO's decisions become the inputs of the search engines' algorithms, and vice versa. By dynamically responding to each other, the substance of trademark applications will change over time, and, ideally, in a way that gradually eliminates "easy" cases. Moerland and Freitas argue that so far, government search engines have not developed a level of sophistication that can replace human examiners. Future work might explore whether this argument holds true for such "easy" cases, or whether it is more applicable to "hard" cases involving novel or ambiguous marks.

Using adversarial machine learning as a model, we can open up new areas of inquiry in addressing situations where a public agency needs to make decisions based on information provided by a private actor. Adversarial machine learning provides a framework for thinking of dynamic government decision-making systems as responding to added noise. Just like adding random pixels to an image stress tests the AI system that powers a self-driving vehicle, policymakers can think about ways to utilize stress tests provided by private actors to better calibrate law, policy, and administrative decision-making. Thus, administrative agencies investing in machine learning tools and, more importantly, adopting theoretical frameworks about dynamic decision-making can empower them to improve over time.

C.     RISK ASSESSMENT IN THE TRADEMARK ECOSYSTEM

Our study suggests that a supply-side study of trademarks should engage further with the search costs associated with post-registration enforcement, as well as the search costs inherent in the entire brand management process. Getting a trademark registered is important, but the post-registration landscape of enforcement is perhaps is even more important. The largest question, perhaps for a future round of research, concerns the impact of AI on the overall trademark litigation ecosystem, i.e., whether or not search costs may have a similar effect on the trademark system like the patent system, where patent trolling and patent pooling have detrimentally affected the marketplace of patent acquisition and enforcement. With millions of existing trademarks spread across a variety of industries, it is simply infeasible to manually look for potential conflicts and deal with them as they arise. Instead, AI-powered tools can consume this tremendous amount of brand-related data, process it, and present it to the brand owner in a way that filters out noise while giving trademark owners a way forward. In sum, by substantially reducing the costs associated with search, these tools also bolster trademark holders' abilities to protect their IP effectively.

One central question that can be raised from this project is similar to questions raised regarding the use of AI in other contexts: will AI transform trademark law altogether? Of course, given the rapid increase in trademarking activity in the past few decades, one can certainly understand the intuitive appeal of employing a greater use of AI. However, as Gangjee notes, "[t]he seductive appeal of the all-seeing algorithm should be resisted," because it faces, at best, a current set of limitations.[273] We believe, like other AI trademark experts, that while AI has the capacity to refine and improve the process of trademark search and registration, at its best, it should serve to complement, rather than replace, human judgment.[274] Of course, it would be unrealistic to predict that AI-driven judgment will somehow diverge widely from human judgment, mainly because AI is normally trained on decision-making data that is generated by humans. As Gangjee notes, "where the data for a machine learning approach is derived from judicial content analysis—past decisions by human tribunals where factors are coded and correlations derived—the algorithm will behave like the human decision maker it is modelled after, warts and all."[275]

In sum, as our paper has suggested, searching for preexisting trademarks is simply the first step in the process of overall brand management. While we conducted an in-depth look at the search process inherent in the trademark process, the platforms we studied also provide brand management services. Such services provide us with a deeper set of variables that may even go beyond the systems of patent pooling and enforcement that we have seen thus far. The same AI and machine learning tools that power their search engines also power their brand management tools, suggesting that further study of brand management and AI may be warranted.

Consider, for example, the rich set of possibilities that stem from providing a preliminary analysis of risk assessments in trademark search. Much of the existing literature that explore the use of AI-driven risk assessments in government decisions focus mainly on actors with enforcement powers in either criminal justice or administrative law. So far in legal, computer science, and policy literatures, discussions on the use of risk assessment in public policy primarily focus on the implications of AI tools on values like fairness, accountability, and transparency.[276] Risk assessment has been the subject of debate in criminal justice, especially, with applications to sentencing,[277] parole decisions,[278] and bail reform.[279] Scholars have also recently focused attention on critical issues like housing and employment, thus extending discussions on fairness in machine learning to include anti-discrimination and equal protection law.[280] These discussions largely center around the legal problems and implications stemming from the use of "black-box" algorithms in decisions.[281] In particular, the scholarly community is deeply engaged with the possibility that algorithms can learn and reinforce human biases in a way that creates inequitable outcomes for marginalized communities.

---

273. *See* Gangjee, *supra* note 6, at 15.
274. *Id.* at 11 (adopting this view and quoting COMPUMARK WHITE PAPER, ARTIFICIAL INTELLIGENCE, HUMAN EXPERTISE: HOW TECHNOLOGY AND TRADEMARK EXPERTS WORK TOGETHER TO MEET TODAY'S IP CHALLENGES 5 (2018) (observing that AI is "intended to complement, not replace, human analysts")).
275. *See* Gangjee, *supra* note 6, at 11.

Our results suggest that a ripe area for future research could be the use of risk assessments in IP law. Arti Rai describes the theoretical potential for the use of machine learning models in patent applications, and critically notes that many of the equity and justice concerns inherent in areas like crime and housing may not apply to IP contexts in the same way.[282] Given that the stakes are quite different, IP may be a good subject to explore and experiment with risk assessments in legal decision-making. This is especially because the government does not bear the same set of enforcement responsibilities in trademark law.

Engstrom et. al. have explored the idea of surveying the use of AI across government administration.[283] They created a typology of different AI use cases in government such as enforcement, regulatory research, and adjudication.[284] Through the exercise, they defined adjudication specifically as, "[t]asks that support formal or informal agency adjudication of benefits or rights," and note that patent and trademark office applications as an example.[285] Our study suggests a general approach and method for assessing the interplay between the government's adjudication system and the private sector, and this general framework could also be applied to other areas of law as well. Zoning, licensure, and social security benefits claims are all examples of the government adjudicating the benefits and rights of private parties, and the ability to assess how AI-driven systems work in these spaces will likely be a rich, new research area.

---

276. *See* Solon Barocas, Moritz Hardt & Arvind Narayanan, FAIRNESS AND MACHINE LEARNING, https://fairmlbook.org/ (last updated Dec. 6, 2019, 3:49 PM).
277. *See* John Monahan & Jennifer L. Skeem, *Risk Assessment in Criminal Sentencing*, 12 ANN. REV. OF CLINICAL PSYCHOL. 489 (2016).
278. *See* Megan Stevenson, *Assessing Risk Assessment in Action*, 103 MINN. L. REV. 303, 304 (2019).
279. *See generally* Jon Kleinberg, Himabindu Lakkaraju, Jure Leskovec, Jens Ludwig & Sendhil Mullainathan, *Human Decisions and Machine Predictions*, 133 Q.J. ECON. 237 (2018).
280. *See generally* Solon Barocas & Andrew D. Selbst, *Big Data's Disparate Impact*, 104 CALIF. L. REV. 671 (2016).
281. *See generally* Andrew D. Selbst & Solon Barocas, *The Intuitive Appeal of Explainable Machines*, 87 Fordham L. Rev. 1085 (2018).
282. *See* Rai, *supra* note 3.
283. *See generally* Engstrom et al., *supra* note 64.
284. *See generally id.*

With respect to trademark law, our work suggests that greater employment of risk assessments can play a central role in brand management after registration. For example, one core question in studying risk assessments in the law is whether legal decisions can be effectively mapped onto mathematical relationships. However, the process by which human decision-makers give effect to legal rules is inherently a black box. The 2(d) "likelihood of confusion" test reflects the way that law typically creates somewhat nebulous rules. These rules only become effective because human beings (judges, bureaucrats, etc.) interpret them and create standards for how they should be applied. Giving explicit written reasons for decisions is one way that decision-makers can communicate how they arrived at decision.[286]

Importantly, we echo Rai's central point that transparency and explicability are not necessarily the same thing in the intellectual property context.[287] Explicability is an elusive goal in these sorts of agency decisions because human decision-making is inherently a black box. Similarly, machine learning models may also suffer from this lack of explicability.[288] In the context of a 2(d) denial of a trademark application, it may be impossible to truly explain how either a trademark officer or a machine learning models making determinations about likelihood of confusion.

However, as we show, not all hope is lost because one need not understand precisely why a potential mark will be rejected as a 2(d) violation in order to make decisions. Simple diagnostic tools can provide insights into how decisions are being made. In our case, we evaluate trademark search engines that deploy AI to power their results and find that they in general reduce search costs for potential users. In doing so, we demonstrate that one way forward in studying risk assessments in the law is to evaluate the outputs of AI models. We specifically focus on search results in trademark search engines, but this general framework could be applied broadly across various domains.

---

285. *Id.* at 10.
286. *See generally* Frederick Schauer, *Giving Reasons*, 47 STAN. L. REV. 633 (1995).
287. *See* Rai, *supra* note 3.
288. *See generally* Selbst et al., *supra* note 281.

Here, the employment of predictive analytics can also help conserve private resources spent on enforcement. To illustrate, at least two search engines use percentage-based scores to assess the risk that these marks, if selected, would cause legal concern.[289] If a firm realizes that a mark with a "very high risk" score has been approved by the USPTO, that information will allow it to prioritize taking legal action against the holder of the conflicting mark, rather than wasting resources pursuing marks that are not especially damaging. Alternatively, a firm that selects a mark with a "very high risk" score faces a high level of vulnerability due to the likelihood of a legal challenge to the mark's selection.

Figures 22 and 23 show what risk assessments look like in the trademark search context. Figure 22 shows an example report from TrademarkNow,[290] and the numerical figures on the left indicate a mark's riskiness of running into a likelihood-of-confusion denial. Figure 23 shows the distribution of risk scores from Markify.[291] A user can use these services to prioritize their search results, and evaluate their registration strategy in light of risk assessment scores. From a user's perspective, one can easily focus on "high" risk results and determine whether to proceed on that basis, while paying less attention to "medium" and "low" risk results. This sort of prioritization is important because the heart of AI-driven trademark search is to reduce the human effort needed to assess likelihood of confusion, and instead focus on other parts of the trademark application process. Because there is a huge supply of potentially conflicting trademarks, the effort required to make a determination about each potential conflict can add up quickly. As we showed earlier, any given searched mark could be expected to return at least ten potential conflicts, and sometimes in excess of two hundred. An AI-generated risk score removes much of this guess work, and would be especially helpful for edge cases. The user can focus on the "high" risk results and tailor their application to avoid conflict with these results. Without wasting time and effort on marks that would be unlikely to cause problems anyway, the user would save a potentially enormous amount of time and costs associated with hiring a trademark attorney.

---

289. Both TrademarkNow and Markify provides these assessments. *See, e.g.*, *Unlimited Trademark Screening & Analysis with ProSearch™*, MARKIFY, https://www.markify.com/services/prosearch-temp.html (last visited Jan. 23, 2021) (discussing its metrics for "statistical risk analysis"). TrademarkNow's product description says that its services allow a user to "a clear picture of risk across all regions of interest in seconds and review your clearance search results ranked and analyzed in order of threat." *Clearance Search – NameCheck™*, TRADEMARKNOW, https://www.trademarknow.com/products/namecheck (last visited Jan. 23, 2021).

290. This image is taken from TrademarkNow's demo page: https://www.trademarknow.com/name-check-video.

291. These are drawn from Markify's Comprehensive Reports rather than the knockout searches we used earlier.

**Figure 22: Sample Risk Scores from TrademarkNow's Platform**

Figure 23: Risk Level Distributions in Markify Dataset

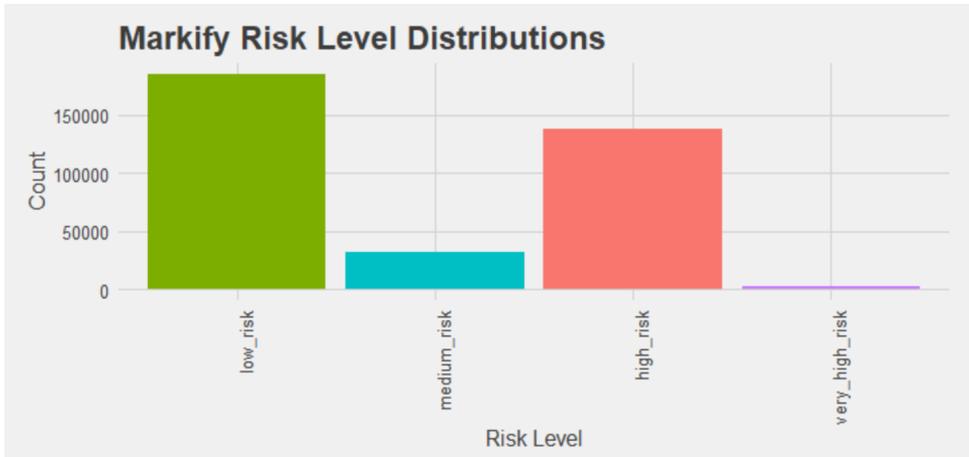

Another way we see the various ways by which underlying AI may work is by looking at how each search engine deals with similarity. Again, we do not know the exact mechanics of how each search engine defines similarity or the thresholds that each chooses when optimizing information retrieval. However, we do have the outputs and can diagnose how well those outputs fit our predefined metrics. In particular, we can use Levenshtein distance[292] to analyze the results produced by each search engine. A Levenshtein distance is calculated between two text strings by looking at the number of edits—additions, subtractions, substitutions, and deletions—that it takes to get from one string to another. Figure 24 shows the distribution of Levenshtein distances across some of our search engines. A quick look at each search engine's distributions shows how their underlying algorithms may prioritize different kinds of results. For instance, Corsearch returns relatively few extremely close matches, likely because its algorithm is more focused on phonetic matching. Trademarkia returns a relatively large number of exact or close matches, indicating that it is more concerned with finding obvious candidates.

---

292. *See generally* Frederic P. Miller & Agnes F. Vandome, DAMERAU-LEVENSHTEIN DISTANCE: INFORMATION THEORY, COMPUTER SCIENCE, VLADMIR LEVENSHTEIN, STRING METRIC, STRING (COMPUTER SCIENCE), TRANSPOSITION (MATHEMATICS) (John McBrewster, Ed., 2010).

Figure 24: Distribution of Levenshtein Distances

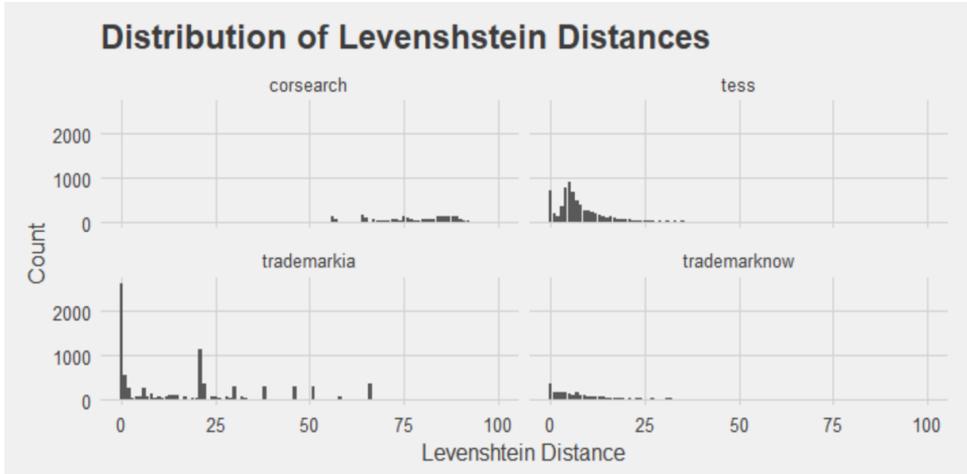

Figure 25: Lollipop Chart of Median Levenshtein Distances

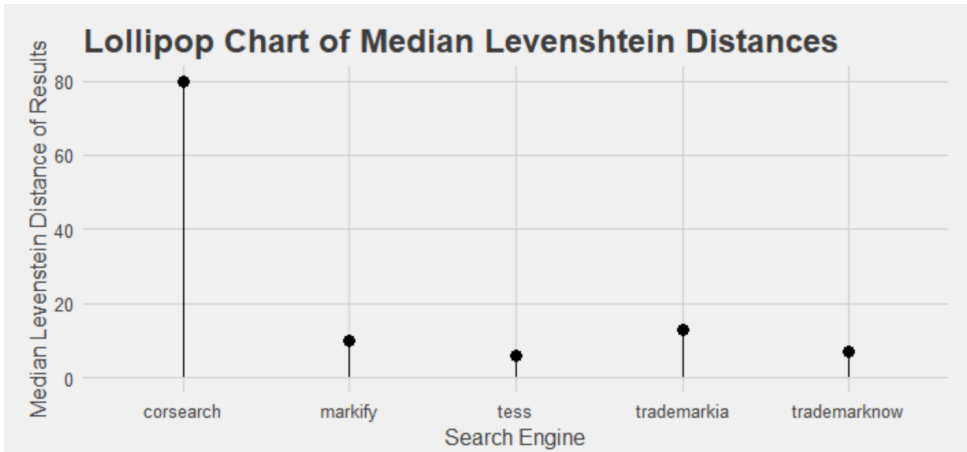

Delving deeper, we can see this relationship even more clearly. Figure 24 shows the median Levenshtein Distance for results, separated by search engine. TESS, Markify, Trademarkia, and TrademarkNow all tend to return results that are fairly close to the searched mark. Corsearch is a clear outlier here, again, because its algorithm is likely prioritizing different kinds of results.

Using these simply defined metrics and plots, we can see how these relatively straightforward tools can be used to understand and diagnose AI systems. In particular, by focusing our attention on the outputs of these search engines, we can perform apples-to-apples comparisons among them to make inferences about how their underlying algorithms work. These inferences can then enrich our general theory of search costs in the trademark spaces because they suggest that firms look for a variety of attributes within their initial search for a trademark. These attributes are likely directly related to the outputs that we uncovered in this study, giving us insight into how users make decisions about which AI tools to use in their searches and brand management efforts.

Here, it is important to note that given the sheer power of predictive analytics coupled with massive amounts of data storage and retrieval, there is at least some potential for AI to surpass human judgment and performance when it comes to analyzing and integrating a much wider array of variables in its assessments.[293] But this may not always be a good thing, particularly where subjective judgment (or survey evidence) is relied upon in court. In some cases, risk assessments can result in a mechanistic, formalistic prediction of liability. Where AI lacks the human ability to consider context, it may result in a higher, expanded prediction of likelihood of confusion.[294] This outcome suggests at first that a greater reliance on AI at the front end in the registration process may actually reduce the incidence of infringement and confusion at the back end (after the mark has entered the market).[295] But this may leave out the consumer in the process of determining actual confusion on the back end. In fact, Dev Gangjee has observed, "[t]he reactions of a real-world consumer, so often alluded to in trademark doctrine, may be muted even further as a result."[296]

There are other concerns raised by an overreliance on AI in risk assessment strategies. Given the large number of marks that are not in use, but which remain registered or may be unregistered, there is also a risk that assessments may not reflect the reality of the existing marketplace. Here, AI-driven tools may not be able to distinguish between marks that are actually in use from those that are just claimed for use (but not actually in use yet), thereby creating a greater risk of false positives for likelihood of confusion.[297] The converse of this is also created by the limited ability of AI to accurately assess other risks beyond infringement. For example, the risk of dilution through blurring or tarnishment or inclusion of common law trademarks in assessments present other risks that produce more false negatives and enable potential free-riding activity.[298]

---

293. *See* Gangjee, *supra* note 6, at 11.
294. *See id.* at 12–13.
295. *See id.* at 13.
296. *See id.*
297. *See id.* at 14.
298. *See id.*

Future studies, of course, could conduct a similar analysis to study these aspects of trademark search engines. One could generate a list of valuable trademarks and run tests on each search engine to determine how well they flag potential conflicts. Again, the framing here is important. Whereas we looked at the economics from the perspective of a registrant, there is also a fascinating world of study to explore from the perspective of a trademark holder, after a trademark has been granted. One core area worth studying further is how AI fits into an emerging divide in trademark law between those who benefit from utilizing an enforcement strategy focused on litigation and those who do not.[299] There may be other ways to generate data surrounding new trademark applications or enforcement strategies, and new experiments could lead to novel new insights.

Last, while our empirical study is limited to basic word search marks, there is room to explore all of the ways that AI is transforming the trademark search space in terms of visual marks and logos, as well. As computer vision tools develop, a follow up study could see how well each search engine returns close visual matches. This sort of study would be fascinating because it would present an interesting exploration of how brands protect elements of their logos and how the USPTO thinks about visual similarity.

D.     FUTURE WORK

Our study opens up several possibilities for future work on trademark search and artificial intelligence. In particular, we have established a reproducible method for searching trademark applications, and evaluating how well various search engines do on various metrics. Other researchers can expand the set of searches, change the metrics, or analyze new data in different ways.

In particular, one possible extension of our work is using pending trademark applications instead of previous applications that got 2(d) citations. One could scrape new trademark applications, search these names in the search engines, and wait to see which ones are rejected by the USPTO. This type of study effectively achieves what we did with previous 2(d) citations, but with the benefit of evaluating marks that have yet to be reviewed by the USPTO.

Otherwise, a further area of study could be examining whether there are differences between different types of registrants. Although we did not use this information in our analysis, trademark applications also have information about the registrant. Examining whether there are differences in applications and 2(d) denial rates among different types of registrants could be interesting. For instance, examining the difference between repeat registrants and first-time registrants, companies in different industries, and various other factors could further enrich our understanding of how trademark search engines work.

---

299. On this point, see generally Glynn Lunney, *Two-Tiered Trademarks*, 56 HOUS. L. REV. 295 (2018).

Finally, we raise questions about the interplay between AI-powered trademark searches and USPTO trademark-granting activity. This area has been explored theoretically in patent literature already, and we expand this discussion to trademarks. While we provide some preliminary evidence about how trademark search engines work, more work should be done to study the interplay directly and how trademarks evolve over time, if at all.

## CONCLUSION

In this paper, we outlined a framework for understanding the economics of trademarks from the perspective of trademark holders, and we examined how AI is rapidly changing the search costs involved with trademark registration and acquisition. We then conducted a novel empirical study that explores how AI is used by trademark search engines, comparing the results from various AI-related private vendors. Our research suggests a greater need for trademark scholars to consider a foundational transformation attributable to AI, where the trademark holder essentially becomes a consumer of trademarks. Such a transformation necessitates a greater attention to the supply of, rather than the demand for, trademarks. Finally, we discussed the implications our findings have for IP law, and the role of AI and search in legal contexts. Going forward, we hope this paper opens up an exploration of the impact that AI will have on trademarks, search costs, and legal administration more broadly.